\documentclass[preprint,authoryear,3p,10pt]{elsarticle}

\makeatletter
\def\ps@pprintTitle{%
 \let\@oddhead\@empty
 \let\@evenhead\@empty
 \def\@oddfoot{\centerline{\thepage}}%
 \let\@evenfoot\@oddfoot}
\makeatother

\usepackage{graphics}
\usepackage{graphicx}
\usepackage{amsmath,amssymb,esint}
\usepackage{lineno}
\usepackage{multirow}
\usepackage{subfig}

\usepackage[colorlinks]{hyperref}

\usepackage[usenames,dvipsnames]{xcolor}
\newcommand{\revone}[1]{\textcolor{black}{#1}}

\journal{Ultrasound in Medicine and Biology}

\begin{document}

\begin{frontmatter}

\title{Modelling lipid-coated microbubbles in focused ultrasound applications at subresonance frequencies}

\author{Jonas G\"ummer}
\author{S\"oren Schenke}
\author{Fabian Denner\corref{cor1}}

\address{Chair of Mechanical Process Engineering, Otto-von-Guericke-Universit\"{a}t Magdeburg,\\ Universit\"atsplatz 2, 39106 Magdeburg, Germany}

\cortext[cor1]{Corresponding Author: Email: fabian.denner@ovgu.de}

\begin{abstract}
We present a computational study of the behaviour of a lipid-coated SonoVue microbubble with initial radius $1 \, \mu \text{m} \leq R_0 \leq 2 \, \mu \text{m}$, excited at frequencies ($200-1500 \, \text{kHz}$) significantly below the linear resonance frequency and pressure amplitudes of up to $1500 \, \text{kPa}$, an excitation regime used in many focused ultrasound applications. The bubble dynamics are simulated using the Rayleigh-Plesset equation and the Gilmore equation, in conjunction with the Marmottant model for the lipid monolayer coating. Also, a new continuously differentiable variant of the Marmottant model is introduced. Below the onset of inertial cavitation, a linear regime is identified in which the maximum pressure at the bubble wall is linearly proportional to the excitation pressure amplitude and, likewise, the mechanical index. This linear regime is bounded by the Blake pressure and, in line with recent {\em in vitro} experiments, the onset of inertial cavitation is found to occur approximately at an excitation pressure amplitude of $130 - 190 \, \text{kPa}$, dependent on the initial bubble size. In the nonlinear regime the maximum pressure at the bubble wall is found to be readily predicted by the maximum bubble radius and both the Rayleigh-Plesset and Gilmore equations are shown to predict the onset of sub- and ultraharmonic frequencies of the acoustic emissions compared to {\em in vitro} experiments. Neither the surface dilatational viscosity of the lipid monolayer nor the compressibility of the liquid have a discernible influence on the studied quantities, yet accounting for the lipid coating is critical for the accurate prediction of the bubble behaviour. The Gilmore equation is shown to be valid for the considered bubbles and excitation regime, and the Rayleigh-Plesset equation also provides accurate qualitative predictions, even though it is outside its range of validity for many of the considered cases.
\end{abstract}

\begin{keyword}
Ultrasound contrast agents \sep Focused ultrasound \sep Nonlinear bubble dynamics \sep Acoustic emissions \sep Inertial cavitation \\~\\
\textcopyright~2021. This manuscript version is made available under the CC-BY-NC-ND 4.0 license. \href{http://creativecommons.org/licenses/by-nc-nd/4.0/}{http://creativecommons.org/licenses/by-nc-nd/4.0/}
\end{keyword}

\end{frontmatter}



\hypersetup{citecolor=black,linkcolor=black}

\section*{Introduction}
\label{sec:intro}

Over the past decade, lipid-coated microbubbles have found a steadily increasing number of diagnostic and therapeutic applications using focused ultrasound \citep{Wan2015, Escoffre2016}, such as  targeted drug delivery \citep{Mulvana2013, Roovers2019} and the treatment of vascular thrombosis \citep{Bader2016}, beside their originally intended use as contrast agents in medical imaging \citep{Tang2011, Christensen-Jeffries2020}.
In focused ultrasound (FUS) treatments utilising microbubbles, ultrasound with large excitation pressure amplitudes of several hundred kPa, at frequencies ranging from several hundred kHz to several MHz, is employed \citep{Escoffre2016}, which leads to strong oscillations of the microbubbles. The acoustic emissions and microstreaming generated by these bubble oscillations promote a physiological response, increasing the permeability of cell membranes \citep{Helfield2016,Qin2018}, breaking cellular structures \citep{Ohl2006,Yuan2015} or causing cell lysis \citep{Chen2003c,Tandiono2012}. Particularly the transient localised opening of the blood-brain barrier for targeted drug delivery using FUS together with coated microbubbles has received considerable research attention \citep{Sun2017, Jones2018}, with {\em in vivo} studies in animals \citep{Tran2020} and first {\em in vivo} trials in humans \citep{Carpentier2016, Lipsman2018} reporting an increased drug uptake and demonstrating significant clinical potential.  

A general concern of FUS, especially at subresonance excitation frequencies ($f_\text{a} < f_\text{res}$) with respect to the linear resonance frequency of the microbubbles ($f_\text{res} \approx 2 - 10 \, \text{MHz}$) in conjunction with the large excitation pressure amplitudes often utilised in FUS, is the generation of acoustic emissions with large pressure peaks. These large pressure peaks lead to large shear stresses and large amounts of heat, which can cause tissue and cell damage \citep{terHaar2011, Izadifar2017}, such as hemolysis as a result of inertial cavitation \citep{Chen2003c}.  
During ultrasound imaging, excitation frequencies close to or above the linear resonance frequency of the microbubbles ($f_\text{a} \gtrsim f_\text{res}$) and small excitation pressure amplitudes ($\Delta p_\text{a} < 100 \, \text{kPa}$) are applied \citep{Qin2009}, and the acoustic emissions likewise have a small amplitude. The large excitation pressure amplitudes ($\Delta p_\text{a} > 100 \, \text{kPa}$) applied in FUS therapies cause inertial cavitation, which leads to the generation of shock waves with significant subharmonic and ultraharmonic content \citep{Song2019}; these are only produced above a frequency-dependent minimum amplitude of the excitation pressure \citep{Helfield2019}. 
Successful FUS treatments, therefore, hinge on techniques to monitor and control key treatment parameters in real time. Such techniques have been in development for the past decade and primarily rely on the frequency spectrum of acoustic emissions, {\em e.g.}~the amplitude of sub- and ultraharmonic emissions, to optimise treatment efficacy and ensure treatment safety \citep{OReilly2012, Sun2017, Jones2018, Gorick2018}.
A detailed knowledge and control of the amplitude and frequency content of the resulting acoustic emission are, therefore, critical for the success and safety of FUS treatments.

With therapeutic FUS applications using lipid-coated microbubbles maturing rapidly through {\em in vitro} and {\em in vivo} studies, the ability to predict and analyse the complex physical phenomena observed during the oscillation and collapse of coated microbubbles using computer models can provide a detailed insight into the physical mechanisms governing the bubble behaviour as well as the resulting cell and tissue manipulation \citep{Helfield2019}. For instance, the pressure and temperature distribution during treatment  are often difficult, and in many instances impossible, to obtain {\em in vitro} or {\em in vivo}, a gap that accurate and robust computer models can fill. 
Yet, such {\em in silico} studies require a comprehensive understanding of the assumptions underpinning the mathematical models and of the ensuing limitations. 

The Rayleigh-Plesset equation \citep{Lauterborn2010} and its extensions to bubbles in compressible liquids \citep{Gilmore1952, Keller1980, Prosperetti1986} are the workhorse of fundamental research on pressure-driven bubble dynamics, and a large number of subtly different Rayleigh-Plesset-based models have been proposed for the modelling of coated microbubbles \citep{Versluis2020}.
The model of \citet{Marmottant2005}, inspired by the area-density-dependent behaviour of phospholipid monolayers \citep{Borden2002, Baoukina2007}, is widely considered to provide the most robust prediction of lipid-coated microbubbles excited with moderate and large pressure amplitudes.
The model accounts for the buckling and rupture of the lipid monolayer based on the bubble radius, whereby the coating exhibits an elastic behaviour only in a limited range of the bubble collapse and expansion, which was shown to be key for the prediction of the resonance response of lipid-coated bubbles at excitation pressure amplitudes $\Delta p_\text{a} \gtrsim 10 \, \text{kPa}$ \citep{Overvelde2010} and of the correct acoustic threshold above which subharmonic and ultraharmonic acoustic emissions are generated \citep{Sijl2010, Paul2010}.

Previous {\em in silico} work and model development has primarily focused on the acoustic regime applied in ultrasonic imaging, with relatively low pressure amplitudes ($< 100 \, \text{kPa}$) and MHz-frequencies. In contrast, in this work we study the response of lipid-coated microbubbles excited with frequencies of $200 - 1500 \, \text{kHz}$ and pressure amplitudes of $10-1500 \, \text{kPa}$, an excitation regime frequently used for focused ultrasound applications. We analyse the influence of the modelling assumptions related to the compressibility of both the liquid and the gas, as well as the influence of the lipid-coating model, on the validity of the governing equations, the onset of inertial cavitation as well as the generation of nonlinear acoustic emissions. To determine the influence of the discontinuous surface tension coefficient resulting from the Marmottant model on the acoustic emissions, which was previously shown to influence the onset of subharmonic emissions \citep{Paul2010}, we also present a new variant of the Marmottant model that yields a continuously differentiable definition of the surface tension coefficient.
The microbubble of choice for this study is a SonoVue bubble (Bracco, Milan, Italy), one of the most comprehensively studied commercially available lipid-coated microbubbles, with an initial radius of $1 \, \mu \text{m} \leq R_0 \leq 2 \, \mu \text{m}$.

\section*{Mathematical models and numerical methods}
\label{sec:numerics}

The response of a single SonoVue bubble in water to a periodic acoustic excitation representative of focused ultrasound applications is simulated. SonoVue bubbles are coated with a phospholipid monolayer \citep{Schneider1999a} and are filled with sulphur hexafluoride (SF$_6$), which has a polytropic exponent of $\kappa = 1.095$ and a density at ambient conditions of $6.17 \, \text{kg/m}^3$. The Rayleigh-Plesset equation and the Gilmore equation are used to model the bubble behaviour, with and without accounting for the rheology of the lipid-monolayer coating, solved using a fifth-order Runge-Kutta method with adaptive time-stepping \citep{Dormand1980}.

\subsection*{Primary equations}

The modified Rayleigh-Plesset equation for large pressure amplitudes, which is regularly used to simulate medical ultrasound applications \citep{Versluis2020}, is considered, which is given as \citep{Marmottant2005}
\begin{equation}
R \ddot{R} + \frac{3}{2} \dot{R}^2 = \frac{p_\text{L} - p_\infty}{\rho_\ell} + \frac{R \, \dot{p}_\text{G}}{\rho_\ell \, c_\ell} ,
\label{eq:modRP}
\end{equation}
where $R$ is the bubble radius, $p_\text{L}$ is the pressure of the liquid at the bubble wall, $p_\infty$ is the pressure of the liquid at infinite distance from the bubble and $p_\text{G}$ is the pressure of the gas inside the bubble, $\rho_\ell$ is the constant density of the liquid and $c_\ell$ is the constant speed of sound of the liquid. The last term on the right-hand side of Eq.~(\ref{eq:modRP}), which is an extension to the classical Rayleigh-Plesset equation, represents the acoustic radiation damping in the liquid \citep{Neppiras1980}. 
For the current study the density and speed of sound of the liquid are assumed to be $\rho_\ell = 1000 \, \text{kg/m}^3$ and $c_\ell = 1476 \, \text{m/s}$, respectively.

In order to account more comprehensively for the compressibility of the liquid, the equation of \citet{Gilmore1952} is considered as an alternative to the Rayleigh-Plesset equation. The Gilmore equation is given as 
\begin{equation}
\left(1-\frac{\dot{R}}{C_\text{L}}\right) R \ddot{R} + \frac{3}{2} \left(1-\frac{\dot{R}}{3 C_\text{L}}\right) \dot{R}^2 =  \left(1+\frac{\dot{R}}{C_\text{L}}\right) H + \left(1-\frac{\dot{R}}{C_\text{L}}\right) R \, \frac{\dot{H}}{C_\text{L}},
\label{eq:gilmore}
\end{equation}
where 
\begin{equation}
C_\text{L} = \sqrt{\Gamma \, \frac{p_\text{L} + B}{\rho_\text{L}}}
\end{equation}
is the liquid speed of sound at the bubble wall and $B = 3.046 \times 10^8 \, \text{Pa}$ is a pressure constant associated with the Tait equation of state that defines, together with the polytropic exponent $\Gamma = 7.15$, the thermodynamic properties of the liquid. The density of the liquid at the bubble wall is
\begin{equation}
\rho_\text{L} = \rho_{\ell} \left(\frac{p_\text{L} + B}{p_0 + B}\right)^{1/\Gamma}.
\end{equation}
The difference between the enthalpy of the liquid at the bubble wall and at infinity, $H$, is defined as
\begin{equation}
H = \frac{\Gamma}{\Gamma - 1} \left(\frac{p_\text{L} + B}{\rho_\text{L}} - \frac{p_\infty + B}{\rho_\infty}\right),
\end{equation}
with
\begin{equation}
\rho_\infty = \rho_{\ell} \left(\frac{p_\infty + B}{p_0 + B}\right)^{1/\Gamma}.
\end{equation}

The pressure of the liquid at infinite distance from the bubble is, including a sinusoidal acoustic excitation, defined as
\begin{equation}
p_\infty = p_0 - \Delta p_\text{a} \sin{(2 \pi f_\text{a} t)},
\end{equation}
where $p_0=10^5 \, \text{Pa}$ is the ambient pressure, $\Delta p_\text{a}$ is the pressure amplitude of the acoustic excitation and $f_\text{a}$ is the frequency of the acoustic excitation.
The gas pressure inside the bubble is
\begin{equation}
p_\text{G} = p_\text{G,0} \left(\frac{R_0^3-h^3}{R^3-h^3} \right)^{\kappa},
\end{equation}
where $R_0$ is the initial bubble radius, $p_\text{G,0}$ is the gas pressure at $R_0$, $\kappa$ is the polytropic exponent of the gas and $h$ is the {\em hard-core} radius of the gas, {\em i.e.}~the radius associated with the van-der-Waals excluded volume. The hard-core radius is estimated based on an effective diameter of SF$_6$ molecules of $d_\text{mol}=550 \, \text{pm}$ and a molecular weight of $m_\text{mol}=146.06 \, \text{g/mol}$.

The influence of the surface tension, the rheology of the lipid-monolayer coating and the viscous dissipation in the liquid are accounted for through the definition of the liquid pressure at the bubble wall, given as \citep{Marmottant2005}
\begin{equation}
p_\text{L} = p_\text{G} - \frac{2 \sigma}{R} - 4 \, \mu_\ell \frac{\dot{R}}{R} - 4 \, \kappa_\text{s} \frac{\dot{R}}{R^2} ,
\end{equation}
where $\sigma$ is the surface tension coefficient, $\mu_\ell = 0.001 \,  \text{Pa s}$ is the dynamic viscosity of the liquid and $\kappa_\text{s}$ is the surface dilatational viscosity of the lipid monolayer. The clean gas-liquid interface has a surface tension coefficient of $\sigma = \sigma_\text{c}$ and a surface dilatational viscosity of $\kappa_\text{s} = 0$. 

Heat transfer is neglected and the compression of the gas in the bubble is assumed to be adiabatic, which is a common assumption for FUS on account of the short timescales of bubble expansion and collapse \citep{Prosperetti1986}. Since mass transfer has a negligible influence for sufficiently large excitation frequencies \citep{Fuster2011}, $f_\text{a} > 100 \, \text{kHz}$, mass transfer is also neglected.

\subsection*{Lipid monolayer model}

For the considered lipid-coated SonoVue microbubble, the surface tension coefficient is given by the model introduced by \citet{Marmottant2005} as
\begin{equation}
\sigma =
\begin{cases}
0 & \text{for} \ R \leq R_\text{buck} \\
\chi \left(\dfrac{R^2}{R_\text{buck}^2} - 1 \right) & \text{for} \ R_\text{buck} < R < R_\text{rupt} \\
\sigma_\text{c} & \text{for} \ R \geq R_\text{rupt}
\end{cases} \label{eq:sigma_marmottant}
\end{equation}
where $\chi$ is the surface elasticity of the lipid monolayer. When the radius of the bubble becomes smaller than \citep{Overvelde2010}
\begin{equation}
R_\text{buck} = \frac{R_0}{\sqrt{1 + \sigma_0/\chi}}, 
\label{eq:Rbuck}
\end{equation}
where $\sigma_0$ is the surface tension coefficient of the lipid-coated bubble at $R=R_0$, the lipid monolayer cannot compress any further and begins to buckle, as a result of which the surface tension effectively vanishes. In contrast, when the bubble expands to a radius larger than 
\begin{equation}
R_\text{rupt} = R_\text{buck} \, \sqrt{1+\frac{\sigma_\text{c}}{\chi}},
\label{eq:Rrupt}
\end{equation} 
the lipid monolayer ruptures and, as a consequence, the clean gas-liquid interface is laid bare. 

\begin{figure}[t]
  \centerline{\includegraphics[scale=0.8]{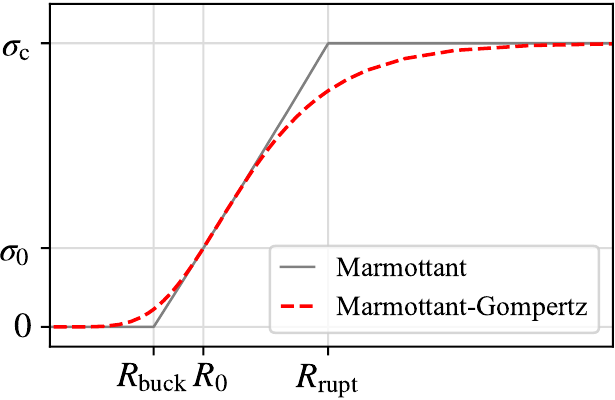}}
\caption{Comparison of the radius-dependent surface tension coefficient for lipid monolayers given by the model of \citet{Marmottant2005}, Eq.~(\ref{eq:sigma_marmottant}), and the Marmottant-Gompertz model, Eq.~(\ref{eq:sigma_gompertz}), for $\sigma_0= 0.020 \, \text{N/m}$, $\sigma_\text{c} = 0.072 \, \text{N/m}$ and $\chi = 0.5 \, \text{N/m}$. The initial radius $R_0$, the buckling radius $R_\text{buck}$, the rupture radius $R_\text{rupt}$, the initial surface tension coefficient $\sigma_0$ and the surface tension coefficient of the clean gas-liquid interface, $\sigma_\text{c}$, are shown as a reference.}
\label{fig:gompertz}
\end{figure}

The radius-dependent surface tension coefficient of the Marmottant model, defined in Eq.~(\ref{eq:sigma_marmottant}), contains two discontinuities at $R=R_\text{buck}$ and $R=R_\text{rupt}$ \citep{Marmottant2005}, where the surface dilatational modulus, $-R^2 \, \partial\sigma/ \partial (R^2)$, of the lipid monolayer is singular. These discontinuities render the Marmottant model sensitive to the applied time-step when numerically solving the primary ordinary differential equation \citep{Versluis2020}. A continuously differentiable form of the Marmottant model is constructed using a Gompertz function of the form $f(x) = a \, \text{e}^{-b \, \text{e}^{-c x}}$, a special case of the generalised logistics function, with the surface tension coefficient defined as
\begin{equation}
\sigma = \sigma_\text{c} \, \text{e}^{-b \, \text{e}^{c (1-R/R_\text{buck})}}, \label{eq:sigma_gompertz}
\end{equation}
with $a = \sigma_\text{c}$ and $x = R/R_\text{buck}-1$, and where the buckling radius $R_\text{buck}$ is given by Eq.~(\ref{eq:Rbuck}). Enforcing $\sigma_0$ for $R_0$, the coefficient $b$ is readily given as
\begin{equation}
b = - \frac{\ln (\sigma_0/\sigma_\text{c})}{\text{e}^{c(1-R_0/R_\text{buck})}}.
\end{equation}
Assuming, additionally, that the maximum slope of the Gompertz function is equal to the derivative of the surface tension coefficient given by the Marmottant model at $R = R_\text{buck} \sqrt{1+\sigma_\text{c}/(2 \chi)}$, the coefficient $c$ follows as
\begin{equation}
c = \frac{2  \chi  \text{e}}{\sigma_\text{c}} \, \sqrt{1+\frac{\sigma_\text{c}}{2 \chi}}.
\end{equation}
Figure \ref{fig:gompertz} shows the Marmottant-Gompertz model alongside the Marmottant model for $\sigma_0= 0.020 \, \text{N/m}$, $\sigma_\text{c} = 0.072 \, \text{N/m}$ and $\chi = 0.5 \, \text{N/m}$, properties that are representative for the bubbles considered in this study. The Marmottant-Gompertz model uses the same set of input parameters ($\sigma_0$, $\sigma_\text{c}$, $\chi$) as the original Marmottant model and reproduces its main features, but with a smooth transition between the surface tension regimes. Therefore, it is particularly suited to determine the influence of the discontinuities of the original Marmottant model. Furthermore, it provides a good qualitative approximation of the elastic behaviour of the lipid monolayer observed in experiments \citep{Segers2018}, with a more rapid change of $\sigma$ near $R_\text{buck}$ than near $R_\text{rupt}$.

\subsection*{Considered bubble properties}

Obtaining reliable and precise input parameters for numerical simulations of lipid-coated microbubbles is difficult, since commercially available contrast agents, such as the considered SonoVue bubbles, are polydisperse \citep{Schneider1999a} and the properties of the lipid monolayer depend, for instance, on the insonation history of the bubbles \citep{Borden2002,Qin2009}. Moreover, associating an observed behaviour or measured quantity with a particular bubble size is complicated by reliably measuring the size of the microbubbles and because it is difficult to isolate individual bubbles in experiments. The bubble properties considered in this study are based on data reported in previous studies and, to this end, we particularly chose properties that allow a direct comparison of the investigated bubble characteristics with experiments.

The presented study considers bubbles with an initial radius in the range $1.0 \, \mu \text{m} \leq R_0 \leq 2.0 \, \mu \text{m}$. This range of initial radii includes the mean radius, $\bar{R}_0 \simeq 1.25 \, \mu \mathrm{m}$ \citep{Schneider1999a, Greis2004}, and the modal radius, $\hat{R}_0 \simeq 1.0 \, \mu \mathrm{m}$ \citep{Greis2004}, of SonoVue bubble populations. It also allows a direct comparison with the  {\em in vitro} experiments reported by \citet{Ilovitsh2018} using a single custom-made lipid-coated microbubble with $R_0 = 1.5 \, \mu \mathrm{m}$, as well as a comparison with the corresponding {\em in vivo} experiments. Furthermore, this size range includes the bubble sizes ($R_0 \lesssim 2.1 \, \mu \text{m}$) for which \citet{Song2019} conducted experiments and analysed the cavitation activity and the associated acoustic emissions of SonoVue bubbles.

The clean gas-liquid interface has a surface tension coefficient of $\sigma_\text{c} = 0.072 \, \text{N/m}$, which is representative of a clean surface of an aqueous liquid, and the equilibrium surface tension coefficient of the lipid-coated bubble is $\sigma_0 = 0.020 \, \text{N/m}$ \citep{Overvelde2010, Katiyar2011a}.
Based on the detailed characterisation of SonoVue bubbles published by \citet{Tu2011}, for the considered initial bubble radii in the range $1.0 \, \mu \text{m} \leq R_0 \leq 2.0 \, \mu \text{m}$ the surface elasticity is assumed to be  $\chi = 0.5 \, \text{N/m}$ and the surface dilatational viscosity is assumed to be in the range  $ 5\times 10^{-9} \, \text{kg/s} \leq \kappa_\text{s} \leq 10^{-8} \, \text{kg/s}$, with a focus on $\kappa_\text{s} = 7.5 \times 10^{-9} \, \text{kg/s}$.

The linear resonance frequencies of the considered lipid-coated SonoVue bubbles are $2950 \, \text{kHz} \leq f_\text{res} \leq 7829 \, \text{kHz}$ \citep{Katiyar2011a}, which are well above the applied range of excitation frequencies, $200  \, \text{kHz} \leq f_\text{a} \leq 1500 \, \text{kHz}$.

\section*{Validity of the primary equations}

\begin{figure}[t]
  \centerline{\includegraphics[scale=1]{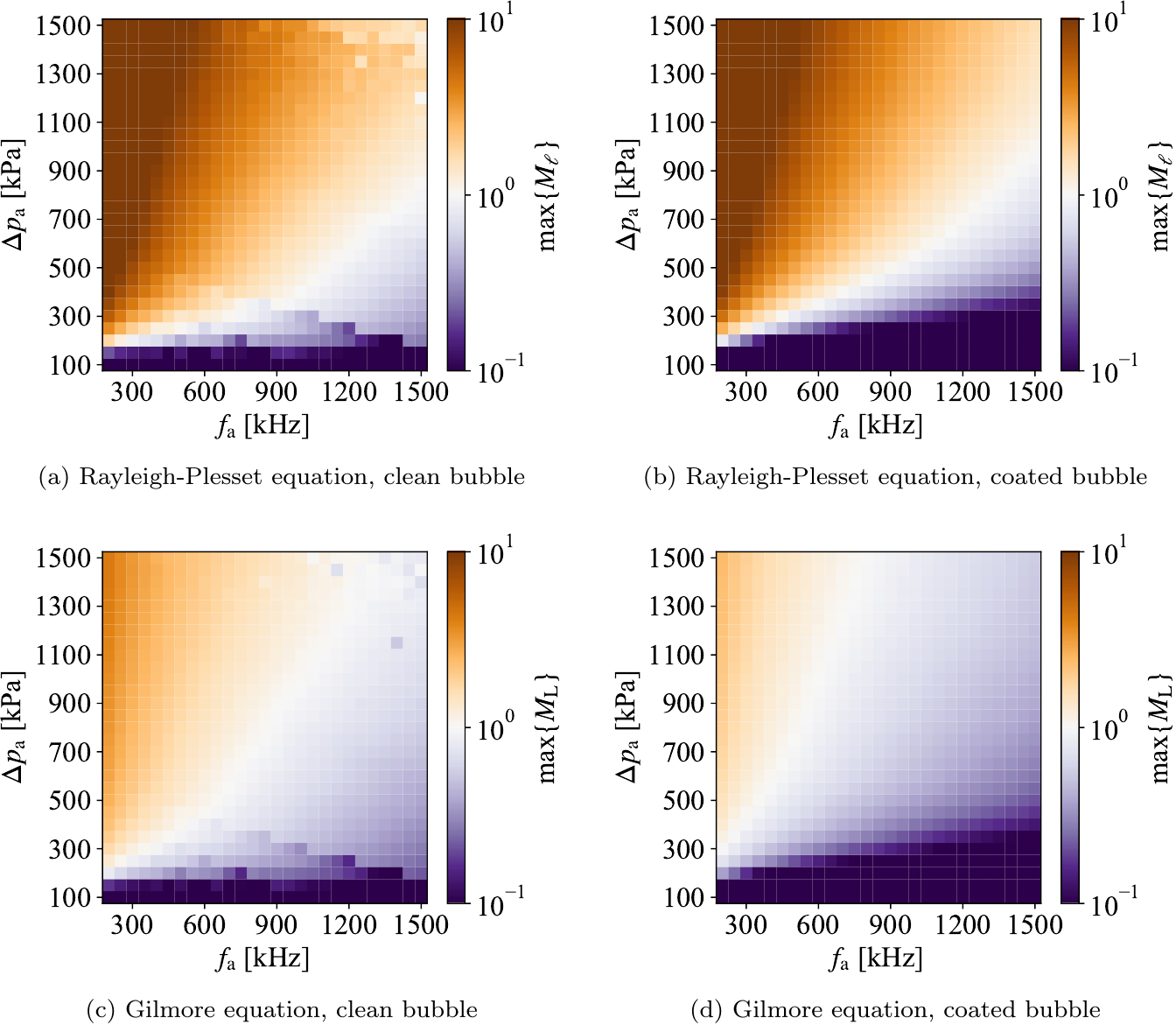}}
\caption{Maximum Mach number of the bubble wall predicted across the considered acoustic excitation regime by the Rayleigh-Plesset equation and the Gilmore equation for a bubble with $R_0 = 1.5 \, \mu \text{m}$ and $h = 310.4 \, \text{nm}$, without and with lipid monolayer coating ($\kappa_\text{s} = 7.5 \times 10^{-9} \, \text{kg/s}$).}
\label{fig:MachLargeAmp}
\end{figure}

\begin{figure}[t]
  \centerline{\includegraphics[scale=1]{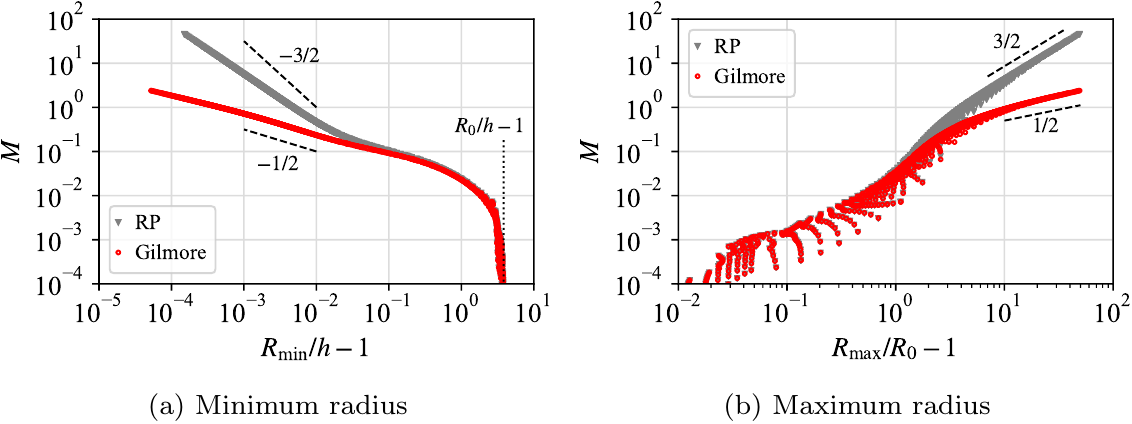}}
\caption{The Mach number of the bubble wall, $M$, as a function of the minimum radius, $R_\text{min}/h-1$, and the maximum radius, $R_\text{max}/R_0-1$, of the coated bubble with $R_0=1.5 \, \mu \text{m}$, $h = 310.4 \, \text{nm}$ and $\kappa_\text{s} = 7.5 \times 10^{-9} \, \text{kg/s}$, predicted by the Rayleigh-Plesset equation (with $M=M_\ell$) and the Gilmore equation (with $M=M_\text{L}$). The theoretical scaling exponents for the Mach number of the wall of an empty cavity with respect to its minimum radius associated with the Rayleigh-Plesset equation ($-3/2$) and the Gilmore equation ($-1/2$) following \citet{Hickling1964} are shown in (a) as a reference. A similar scaling is also evident with respect to the maximum radius in (b).}
\label{fig:R-M}
\end{figure}

In the acoustic excitation regime widely applied in FUS applications, the main difference in assumptions underpinning the Rayleigh-Plesset and Gilmore equations pertains to the compressibility of the surrounding liquid.

The Rayleigh-Plesset equation formulated as presented in Eq.~(\ref{eq:modRP}), including the acoustic radiation term on the right-hand side, follows directly from the Keller-Miksis equation \citep{Keller1980}, 
\begin{equation}
\left(1 - \frac{\dot{R}}{c_\ell}\right) R \ddot{R} + \frac{3}{2} \left(1 - \frac{\dot{R}}{3\, c_\ell}\right) \dot{R}^2 =   \left(1 + \frac{\dot{R}}{c_\ell}\right) \frac{p_\text{G} - p_\infty}{\rho_\ell} + \frac{p_\text{L} - p_\text{G}}{\rho_\ell}  + \frac{R \, \dot{p}_\text{G}}{\rho_\ell \, c_\ell} ,
\label{eq:keller}
\end{equation}
by assuming that the Mach number of the bubble wall is small, $M_\ell  = \dot{R}/c_\ell \ll 1$.
Since the liquid is assumed to be incompressible, the  Rayleigh-Plesset equation (\ref{eq:modRP}) incurs an error proportional to $M_\ell$ and is, thus, only valid for small Mach numbers $M_\ell  \ll 1$.
Nevertheless, feasible results have frequently been obtained with Eq.~(\ref{eq:modRP}) for Mach numbers $M_\ell \sim 1$ \citep{Brenner2002}. With respect to coated microbubbles, numerical results suggest that the acoustic radiation in the compressible liquid surrounding the bubble has a dominant influence on the bubble behaviour for excitation pressures of $\Delta p_\text{a} \approx 100 \, \text{kPa}$ and above \citep{Sojahrood2020}.
The Gilmore equation (\ref{eq:gilmore}) is founded on a variable speed of sound of the liquid and directly incorporates the liquid enthalpy instead of the liquid pressure. In addition to the terms that are first order in the Mach number of the bubble wall, $M_\text{L} = \dot{R}/C_\text{L}$, it also includes terms second order in $M_\text{L}$, but it is not strictly second-order accurate \citep{Prosperetti1986}.
\citet{Gilmore1952} proposed an accurate prediction of the bubble behaviour by Eq.~(\ref{eq:gilmore}) for Mach numbers $M_\text{L} \lesssim 2.2$, while \citet{Hickling1964} even found reliable results using the Gilmore equation for Mach numbers of up to $M_\text{L} \approx 5$. \citet{Prosperetti1986} attributed the success and robustness of the Gilmore equation to the direct use of the liquid enthalpy.

The Mach number of the bubble wall predicted by the Rayleigh-Plesset equation, $M_\ell$, and the Gilmore equation, $M_\text{L}$, across the considered acoustic excitation regime is shown in Figure \ref{fig:MachLargeAmp} for a bubble with $R_0=1.5 \, \mu \text{m}$ and $h = 310.4 \, \text{nm}$, with ($\kappa_\text{s} = 7.5 \times 10^{-9} \, \text{kg/s}$) and without the lipid coating. 
For both primary equations, accounting for the lipid coating reduces the Mach number of the bubble wall.
Nevertheless, even for the lipid-coated bubble, the Rayleigh-Plesset equation predicts a supersonic bubble wall motion ($M_\ell > 1$) for a large part of the considered acoustic excitation regime, with maximum values of $M_\ell > 10$. The Rayleigh-Plesset equation is, therefore, far outside its range of validity. This should be kept in mind for the subsequent analysis. 
The Gilmore equation generally predicts a lower Mach number of the bubble wall motion, as it accounts for the change in density and in speed of sound of the liquid resulting from the considerable change in pressure, with $M_\text{L} < 2$ for all considered cases. The Gilmore equation, thus, remains within its generally accepted range of validity, $M_\text{L} \leq 2.2$, for the considered acoustic excitation regime.

In order to determine when the liquid compressibility starts to have an influence, the Mach number of the bubble wall is shown in Figure \ref{fig:R-M} as a function of the minimum radius and the maximum radius of the coated bubble, for all 1470 combinations of excitation frequency ($200 \, \text{kHz} \leq f_\text{a} \leq 1500 \, \text{kHz}$) and excitation pressure amplitude ($10 \, \text{kPa} \leq \Delta p_\text{a} \leq 1500 \, \text{kPa}$) considered in this study. Overall, there is a strong correlation between the Mach number of the bubble wall and the minimum and maximum bubble radii. As generally expected, the Rayleigh-Plesset equation and the Gilmore equation produce virtually identical results for small Mach numbers, $M < 0.1$. For $M>0.1$, however, the liquid compressibility can no longer be neglected and, as a consequence, the predictions of the two equations increasingly depart from each other. 
As described in detail by \citet{Hickling1964}, the Mach number of the wall of an empty cavity scales theoretically with $M_\ell \propto R^{-3/2}$ in the case of the Rayleigh-Plesset equation and $M_\text{L} \propto R^{-1/2}$ in the case of the Gilmore equation.
The scaling of the Mach number predicted by the Rayleigh-Plesset equation and the Gilmore equation indeed almost reaches these theoretical values for $R_\text{min} <1.01h$, see Figure \ref{fig:R-M}a, as the Mach number of the bubble wall approaches unity and the bubble collapse is inertia-dominated, although the slope is slightly smaller, which is to be expected for a gas-filled bubble. A similar scaling is also evident with respect to the maximum radius in Figure \ref{fig:R-M}b. Neither the initial bubble radius nor the lipid monolayer coating have an appreciable influence on this relationship between bubble size and Mach number.

\section*{Onset of inertial cavitation}

\begin{figure}[t]
  \centerline{\includegraphics[scale=1]{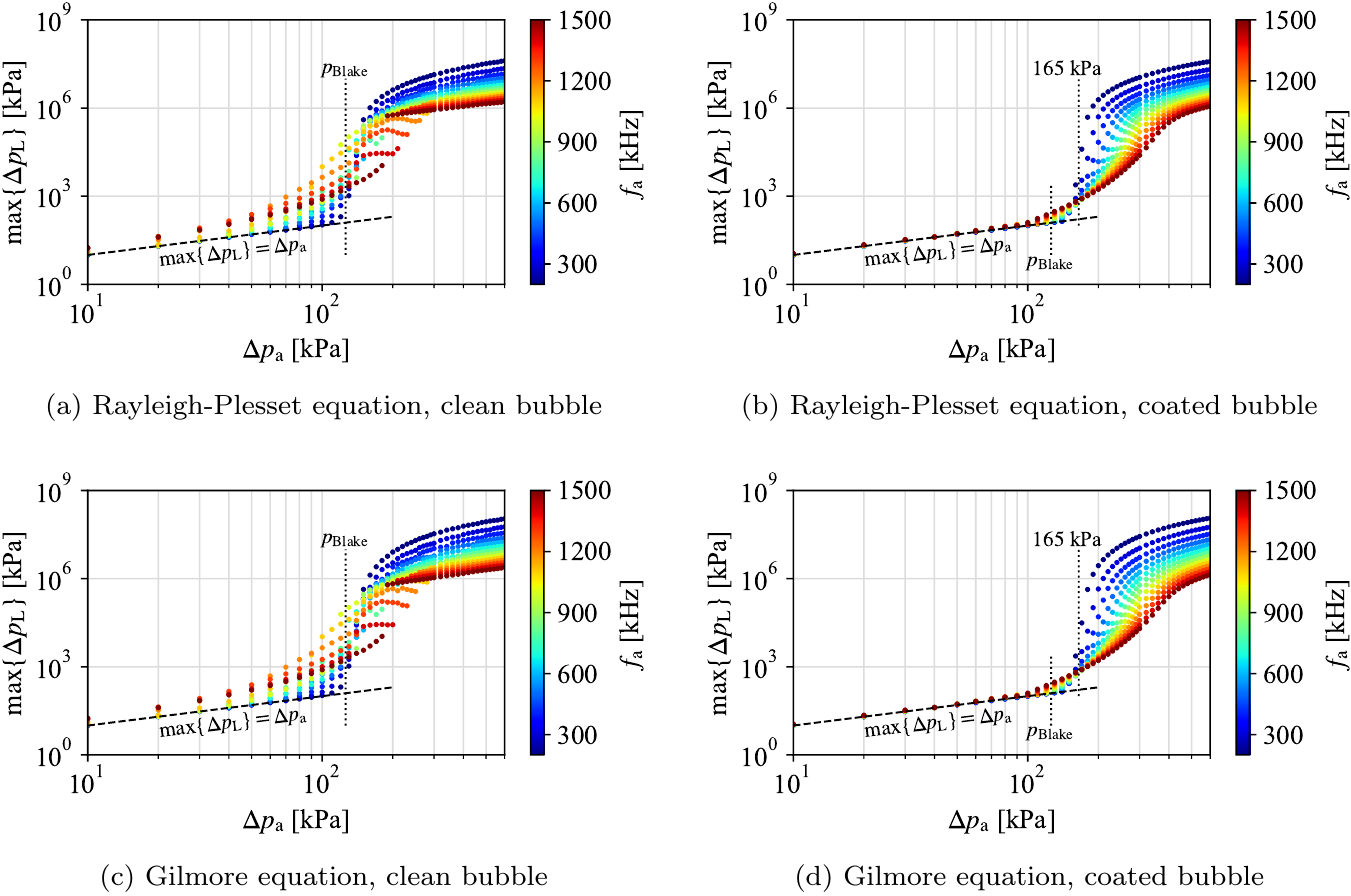}}
\caption{The maximum liquid pressure at the bubble wall, $\Delta p_\text{L} = p_\text{L} - p_0$, of a clean bubble and a coated bubble ($\kappa_\text{s} = 7.5 \times 10^{-9} \, \text{kg/s}$) with $R_0= 1.5  \, \mu \text{m}$ and $h = 310.4 \, \text{nm}$ as a function of the excitation pressure amplitude $\Delta p_\text{a}$, predicted by the Rayleigh-Plesset equation and the Gilmore equation. The colour of the data points represents the excitation frequency $f_\text{a}$. The excitation pressure amplitudes $\Delta p_\text{a} = p_\text{Blake}$ and $\Delta p_\text{a} = 165 \, \text{kPa}$, as well as a linear increase in pressure amplitude, with $\max\{\Delta p_\text{L}\} = \Delta p_\text{a}$, are shown as a reference.}
\label{fig:transition}
\end{figure}

\begin{figure}[t]
  \centerline{\includegraphics[scale=1]{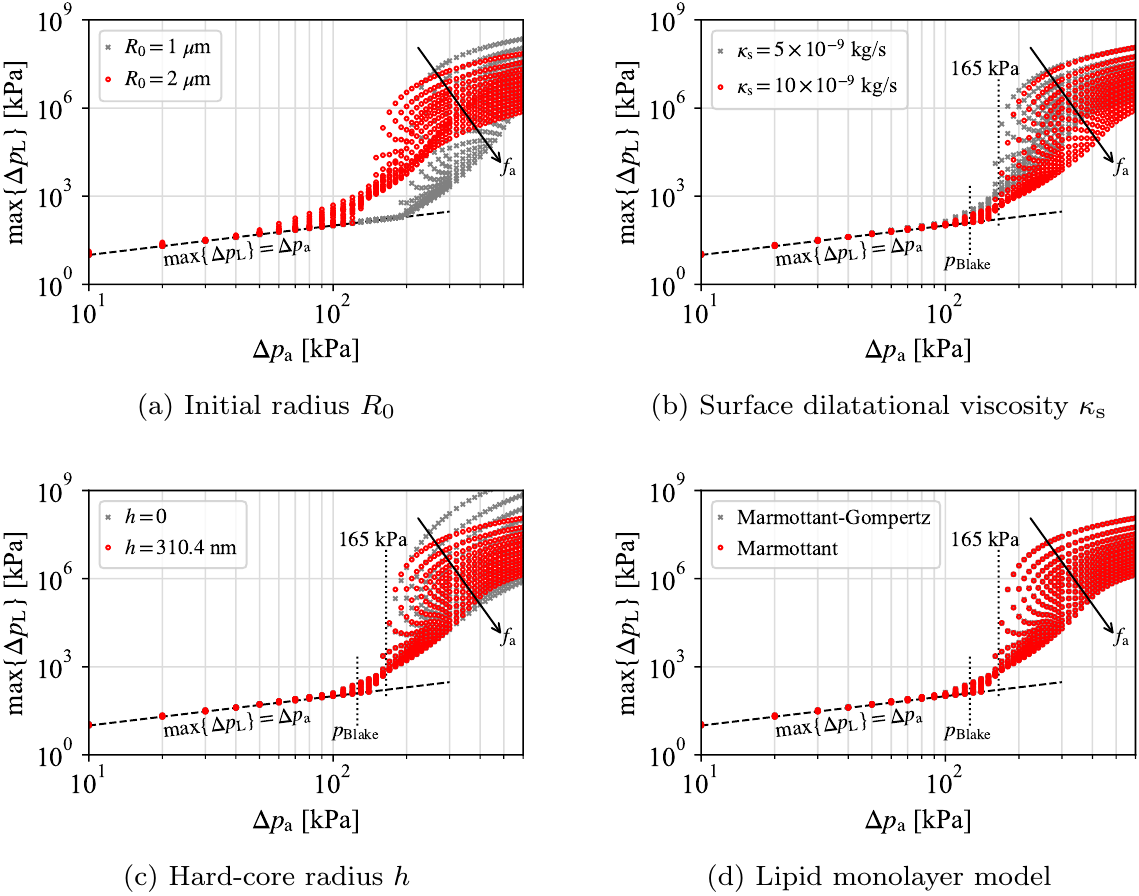}}
\caption{The maximum liquid pressure at the bubble wall, $\Delta p_\text{L} = p_\text{L} - p_0$, of coated bubbles with (a) different initial radii $R_0$, (b) different surface dilatational viscosities $\kappa_\text{s} \in \{5,10\} \times 10^{-9} \, \text{kg/s}$, (c) different hard-core radii $h$ and (d) different lipid monolayer models, as a function of the excitation pressure amplitude $\Delta p_\text{a}$ predicted by the Gilmore equation. If not stated otherwise, the base properties are $R_0 = 1.5 \, \mu \text{m}$ and $\kappa_\text{s} = 7.5 \times 10^{-9} \, \text{kg/s}$, the hard-core radius $h$ is determined based on the properties of SF$_6$, and the Marmottant model is applied. The direction of increasing excitation frequency $f_\text{a}$ is indicated by an arrow. The excitation pressure amplitudes $\Delta p_\text{a} = p_\text{Blake}$ and $\Delta p_\text{a} = 165 \, \text{kPa}$, as well as a linear increase in pressure amplitude, with $\max\{\Delta p_\text{L}\} = \Delta p_\text{a}$, are shown as a reference.}
\label{fig:transitionComp}
\end{figure}

The transition from stable cavitation, with an oscillatory behaviour of the bubble, to inertial cavitation, whereby the bubble collapses strongly and produces a sharp pressure peak, is of significant importance in biomedical applications. Previous studies showed that inertial cavitation may lead to lesions and tissue damage \citep{Ilovitsh2018} as well as hemolysis \citep{Chen2003c}, and may cause changes to the permeability of cell membranes and cell lysis \citep{Coussios2008}. Furthermore, inertial cavitation is often associated with bubble fragmentation \citep{Shi2000, Chomas2001}, which leads to a change in behaviour or dissolution of the fragmented bubbles, and which can be utilised in targeted drug delivery \citep{Wang2014a}.

Recently published {\em in vitro} experiments by \citet{Ilovitsh2018} using ultra high-speed imaging ($\approx 35$ million frames per second) of a lipid-coated microbubble with $R_0 = 1.5 \, \mu \text{m}$ excited at $f_\text{a} = 250 \, \text{kHz}$ suggest the onset of inertial cavitation to occur for excitation pressure amplitudes just below  $165 \, \text{kPa}$, with the bubble in the experiment fragmenting at $165 \, \text{kPa}$ excitation pressure amplitude. Concurrent {\em in vivo} experiments of microbubble-assisted blood-brain-barrier opening with $f_\text{a} = 250 \, \text{kHz}$ in mice showed lesions and acute neurological damage that are typical for inertial cavitation for $\Delta p_\text{a} \geq 150 \, \text{kPa}$ \citep{Ilovitsh2018}.

Figure \ref{fig:transition} shows the maximum liquid pressure generated at the bubble wall, $\Delta p_\text{L} = p_\text{L} - p_0$, of a clean bubble and a coated bubble ($\kappa_\text{s} = 7.5 \times 10^{-9} \, \text{kg/s}$) with $R_0 = 1.5 \, \mu \text{m}$ and $h = 310.4 \, \text{nm}$ as a function of the excitation pressure amplitude $\Delta p_\text{a}$, predicted by the Rayleigh-Plesset equation and the Gilmore equation. For the clean bubble, shown in Figures \ref{fig:transition}a and \ref{fig:transition}c, a change in bubble response with respect to the peak liquid pressure can be observed for all frequencies when the excitation pressure amplitude exceeds the corresponding {\em Blake pressure} \citep{Lauterborn1976}
\begin{equation}
p_\text{Blake} \simeq p_0 + \frac{4 \, \sigma_\text{c}}{3 \, \sqrt{3} \, R_0} \left[1+ \frac{R_0 \, p_0}{2 \, \sigma_\text{c}} \right]^{-1/2}, 
\label{eq:pblake}
\end{equation}
assuming the bubble is filled with noncondensable gas only.
By means of theoretical analysis, \citet{Ilovitsh2018} found the Blake pressure to be a reliable indicator for the onset of inertial cavitation. Below the Blake pressure, higher excitation frequencies yield generally larger pressure amplitudes, $\Delta p_\text{L}$, whereas the pressure amplitude is $\max \{\Delta p_\text{L}\} \simeq \Delta p_\text{a}$ for the smallest considered excitation frequency of $f_\text{a}=200 \, \text{kHz}$ almost up until $\Delta p_\text{a}=p_\text{Blake}$. Above the Blake pressure, smaller excitation frequencies lead to an overall stronger collapse of the bubble.

In the case of the coated bubble, shown in Figures \ref{fig:transition}b and \ref{fig:transition}d, a clear and sudden shift of the  maximum pressure generated in the liquid at the bubble wall can be observed for $\Delta p_\text{a} \approx 150 \, \text{kPa}$. This pressure threshold for the onset of inertial cavitation lies between the Blake pressure as the lower limit, given by the clean bubble, and a pressure amplitude of $165 \, \text{kPa}$, which is the excitation amplitude at which a similar bubble fragmented in the experiments of \citet{Ilovitsh2018}. Below $\Delta p_\text{a} \approx 150 \, \text{kPa}$, the coated bubble responds linearly to the excitation, with $\max \{\Delta p_\text{L} \} \simeq \Delta p_\text{a}$, and the pressure in the liquid at the bubble wall is almost independent of the excitation frequency. 
However, when the excitation pressure amplitude exceeds $\Delta p_\text{a} \approx 150 \, \text{kPa}$, the bubble response becomes nonlinear, with a strong dependence on the excitation frequency $f_\text{a}$. The peak pressure in the liquid exhibits a sudden shift in slope, especially for small excitation frequencies, marking the onset of inertial cavitation. 

The compressibility of the liquid, through the choice of the primary equation, has evidently no appreciable influence on the onset of inertial cavitation, as observed in Figure \ref{fig:transition}, since this transition occurs at Mach numbers of the bubble wall smaller than $0.1$. The maximum pressure generated in the liquid at the bubble wall is shown for different modelling assumptions in Figure \ref{fig:transitionComp}, simulated using the Gilmore equation. The initial bubble size, Figure \ref{fig:transitionComp}a, has a clear influence on the onset of inertial cavitation, with a larger threshold excitation pressure amplitude for smaller bubbles. The onset of inertial cavitation is clearer defined for smaller bubbles, with $\max \{\Delta p_\text{L}\} \simeq \Delta p_\text{a}$ up until the threshold excitation amplitude for all excitation frequencies in the case of $R_0 = 1 \, \mu \text{m}$. The surface dilatational viscosity $\kappa_\text{s}$, Figure \ref{fig:transitionComp}b, the hard-core radius $h$, Figure \ref{fig:transitionComp}c, and the considered lipid monolayer models, Figure \ref{fig:transitionComp}d, have a small or negligible influence on the excitation pressure amplitude at which the onset of inertial cavitation is observed.

\section*{Amplitude of the acoustic emissions}

\begin{figure}[t]
  \centerline{\includegraphics[scale=1]{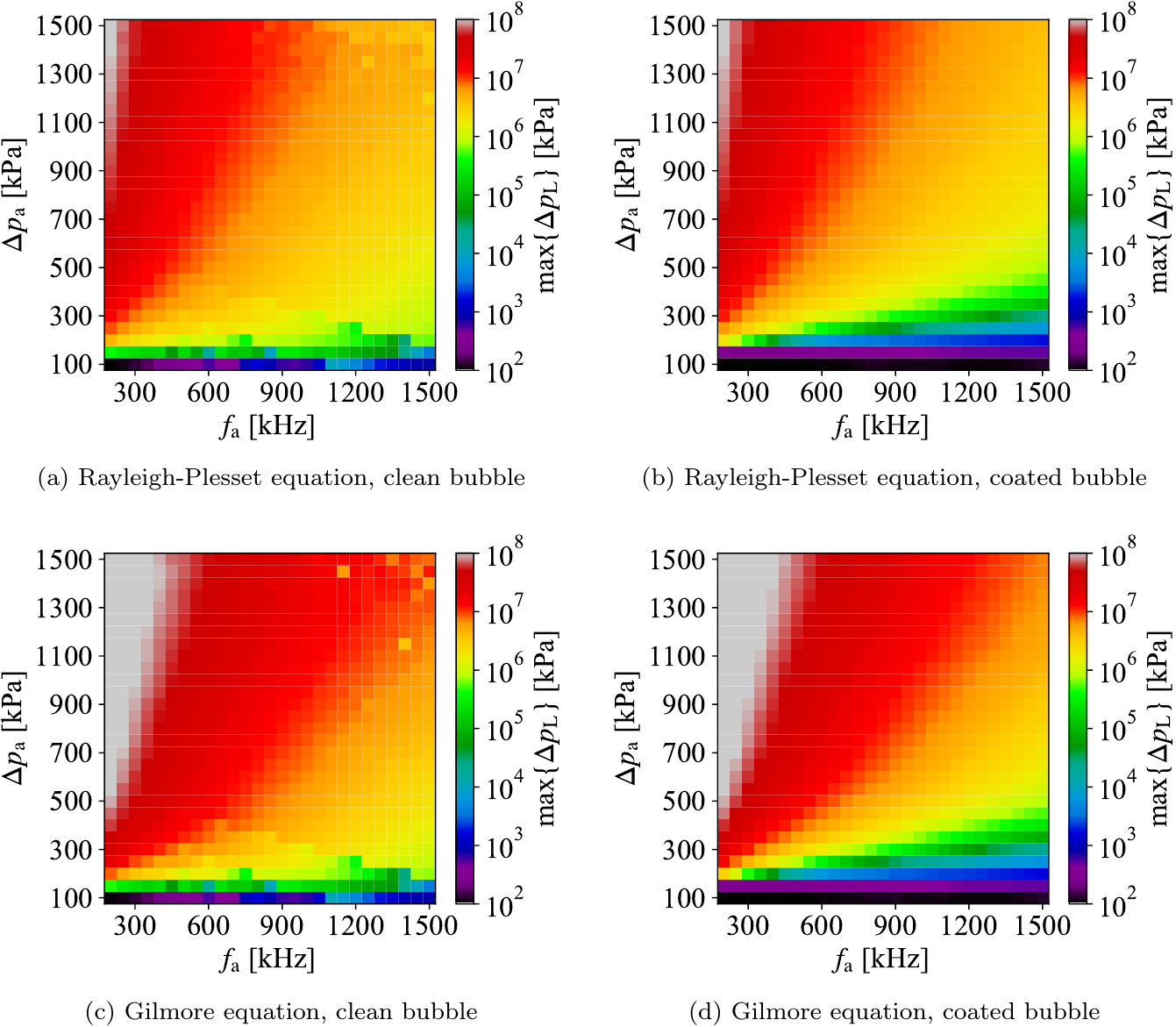}}
\caption{Maximum pressure amplitude generated in the liquid at the bubble wall, $\Delta p_\text{L} = p_\text{L} - p_0$, predicted across the considered acoustic excitation regime by the Rayleigh-Plesset equation and the Gilmore equation for a bubble with $R_0 = 1.5 \, \mu \text{m}$ and $h = 310.4 \, \text{nm}$, without and with lipid monolayer coating ($\kappa_\text{s} = 7.5 \times 10^{-9} \, \text{kg/s}$).}
\label{fig:liqPressureLargeAmp}
\end{figure}

The maximum liquid pressure at the bubble wall is shown in Figure \ref{fig:liqPressureLargeAmp} for the Rayleigh-Plesset equation and the Gilmore equation, for a clean bubble and a coated bubble ($\kappa_\text{s} = 7.5 \times 10^{-9} \, \text{kg/s}$) with $R_0 = 1.5 \, \mu \text{m}$ and $h = 310.4 \, \text{nm}$. The largest pressure is generated by low excitation frequencies and large excitation pressure amplitudes, with peak values ranging from $1.02 \times 10^8 \, \text{kPa}$ to $3.24 \times 10^8 \, \text{kPa}$. At small excitation pressure amplitudes and large excitation frequencies, the coated bubble produces smaller pressure peaks than the clean bubble, irrespective of the modelling assumptions related to the compressibility of the liquid ({\em i.e.}~choice of primary equation). As the excitation pressure amplitude increases and the excitation frequency decreases, the influence of the lipid monolayer coating diminishes. 
The differences between the pressure predictions produced by the Rayleigh-Plesset equation and the Gilmore equation are largest at small excitation frequencies and large excitation pressure amplitudes, which corresponds to large maximum Mach numbers of the bubble wall, see Figure \ref{fig:R-M}.

\begin{figure}[t]
  \centerline{\includegraphics[scale=1]{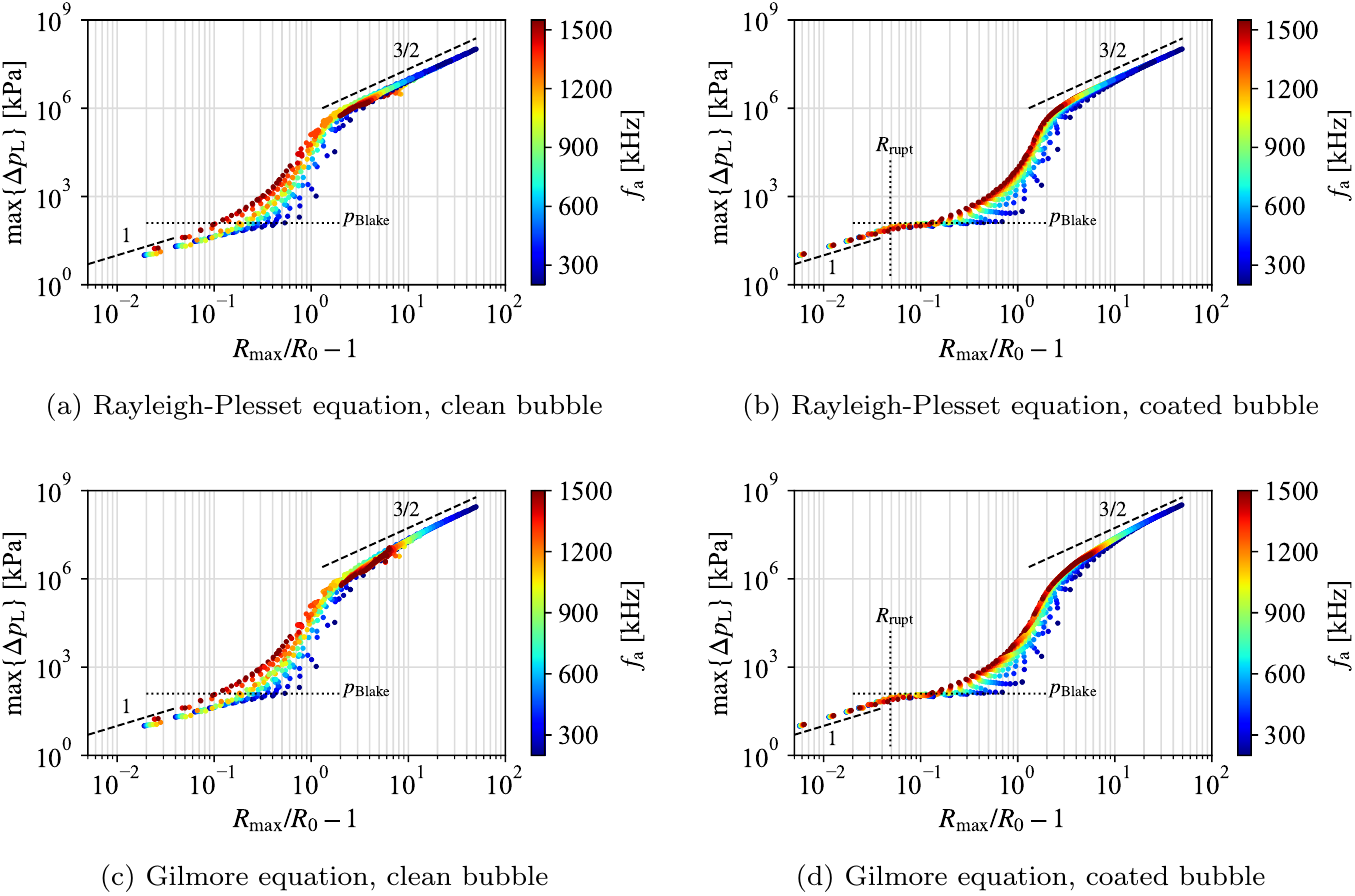}}
\caption{The maximum liquid pressure at the bubble wall, $\Delta p_\text{L} = p_\text{L} - p_0$, as a function of the maximum bubble radius, $R_\text{max}/R_0-1$, predicted across the considered acoustic excitation regime by the Rayleigh-Plesset equation and the Gilmore equation for a bubble with $R_0 = 1.5 \, \mu \text{m}$ and $h = 310.4 \, \text{nm}$, without and with lipid monolayer coating ($\kappa_\text{s} = 7.5 \times 10^{-9} \, \text{kg/s}$). The colour of the data points represents the excitation frequency $f_\text{a}$. The scaling exponent of $1$ represents a linear relationship between the maximum pressure amplitude in the liquid at the bubble wall and the maximum bubble radius. A scaling proportional to $3/2$, the rupture radius  $R_\text{rupt}$ of the liquid monolayer, Eq.~(\ref{eq:Rrupt}), and the Blake pressure $p_\text{Blake}$, Eq.~(\ref{eq:pblake}), are shown as a reference.}
\label{fig:Rmax-pL}
\end{figure}

Figure \ref{fig:Rmax-pL} shows the maximum pressure amplitude in the liquid at the bubble wall as a function of the maximum bubble radius predicted by the Rayleigh-Plesset equation and the Gilmore equation, for all 1470 combinations of excitation frequencies ($200 \, \text{kHz} \leq f_\text{a} \leq 1500 \, \text{kHz}$) and excitation pressure amplitudes ($10 \, \text{kPa} \leq \Delta p_\text{a} \leq 1500 \, \text{kPa}$) considered in this study. While the results exhibit some dependency on the excitation frequency and pressure amplitude, the maximum liquid pressure at the bubble wall associated with the maximum radius of the bubble behaves in a rather predictable fashion. 
In the case of the coated bubble, the liquid pressure at the bubble wall is proportional to $(R_\text{max}/R_0 - 1)$ for radii $R_\text{max}< R_\text{rupt}$, illustrated clearly by the linear pressure increase in Figure \ref{fig:Rmax-pL} with increasing maximum bubble radii for $R_\text{max}<  R_\text{rupt}$. When $R_\text{max} > R_\text{rupt}$, the lipid monolayer is ruptured and, for increasing $\Delta p_\text{a}$, the liquid pressure at the bubble wall briefly stagnates; interestingly this stagnation occurs at $\max\{\Delta p_\text{L}\} \approx p_\text{Blake}$. For the clean bubble, the maximum amplitude of the pressure in the liquid at the bubble wall also increases approximately linearly with increasing maximum bubble radius, up until the peak pressure amplitude generated in the liquid reaches the Blake pressure, $p_\text{Blake}$. However, this linear relationship between $\max\{\Delta p_\text{L}\}$ and $R_\text{max}$ is not as distinct for the clean bubble as it is for the coated bubble. 

\begin{figure}[t]
  \centerline{\includegraphics[scale=1]{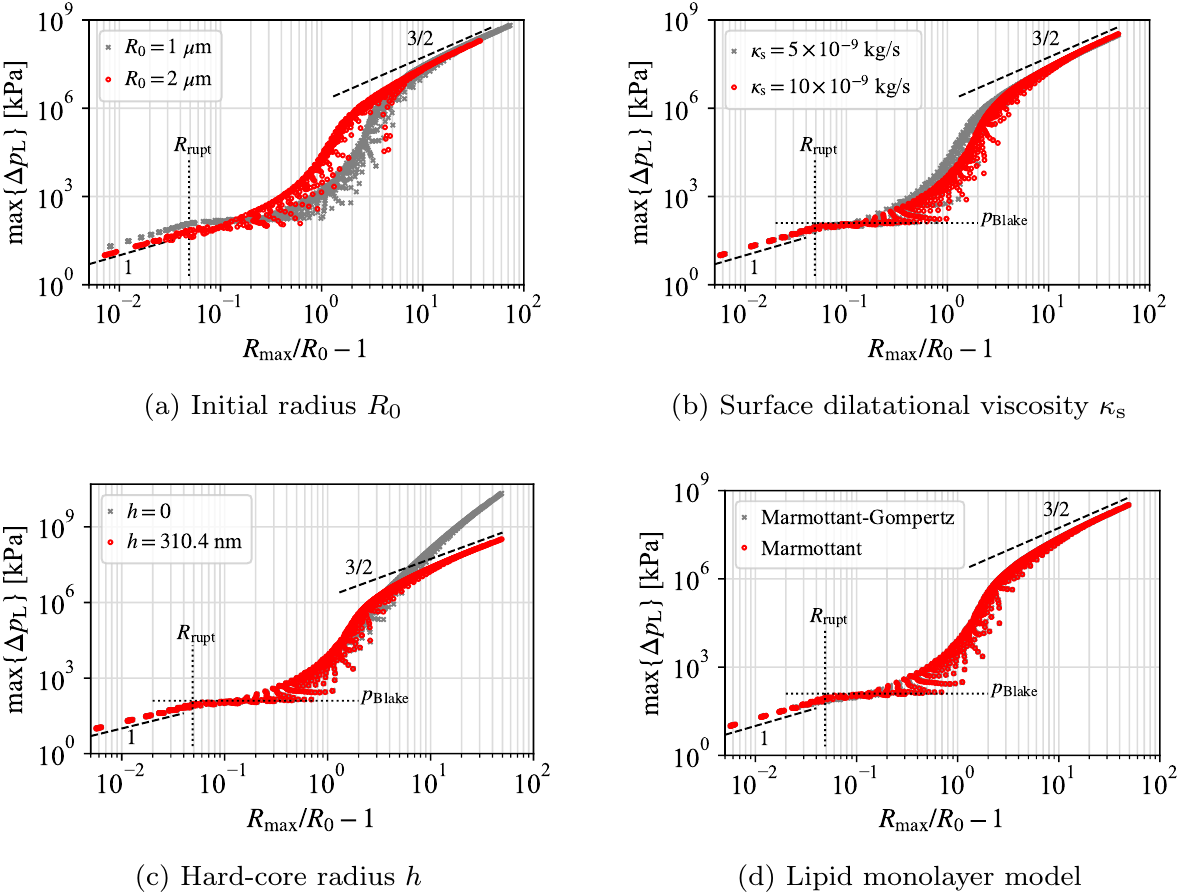}}
\caption{The maximum liquid pressure at the bubble wall, $\Delta p_\text{L} = p_\text{L} - p_0$, as a function of the maximum bubble radius, $R_\text{max}/R_0-1$, of coated bubbles with (a) different initial radii $R_0$, (b) different surface dilatational viscosities $\kappa_\text{s} \in \{5,10\} \times 10^{-9} \, \text{kg/s}$, (c) different hard-core radii $h$ and (d) different lipid monolayer models, predicted by the Gilmore equation. If not stated otherwise, the base properties are $R_0 = 1.5 \, \mu \text{m}$ and $\kappa_\text{s} = 7.5 \times 10^{-9} \, \text{kg/s}$, the hard-core radius $h$ is determined based on the properties of SF$_6$, and the Marmottant model is applied. The scaling exponent of $1$ represents a linear relationship between the maximum pressure amplitude in the liquid at the bubble wall and the maximum bubble radius. A scaling proportional to $3/2$, the rupture radius  $R_\text{rupt}$ of the liquid monolayer, Eq.~(\ref{eq:Rrupt}), and the Blake pressure $p_\text{Blake}$, Eq.~(\ref{eq:pblake}), are shown as a reference.}
\label{fig:Rmax-pL_compProp}
\end{figure}

If the maximum bubble radius becomes significantly larger than $1.2 R_0 \lesssim R_\text{max} \lesssim 2 R_0$, dependent on the excitation frequency, the maximum liquid pressure at the bubble wall increases rapidly for both the clean and the coated bubble, indicating the onset of inertial cavitation. When the maximum bubble radius reaches $R_\text{max} \approx 3 R_0$, the behaviour changes again and the predictions of the Rayleigh-Plesset equation and the Gilmore equation depart from each other. A bubble with this maximum radius collapses with a bubble wall Mach number of $M>0.1$, as observed in Figure \ref{fig:R-M}, and the compressibility of the liquid is no longer negligible. 
For increasing maximum radius $R_\text{max} \gtrsim 3 R_0$, the influence of the lipid monolayer coating reduces and the maximum pressure generated in the liquid asymptotically approaches the scaling $\max\{\Delta p_\text{L}\} \propto (R_\text{max} / R_0-1)^{3/2}$, as observed in Figure \ref{fig:Rmax-pL}, for both the Rayleigh-Plesset equation and the Gilmore equation.

\begin{figure}[t]
  \centerline{\includegraphics[scale=1]{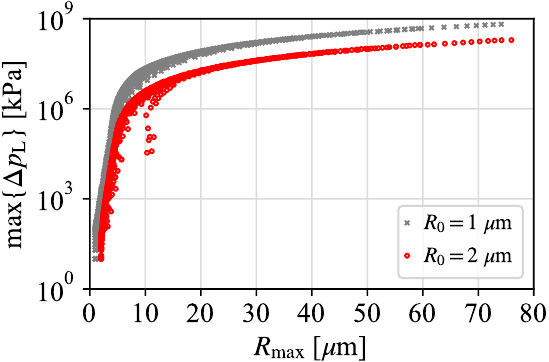}}
\caption{The maximum liquid pressure at the bubble wall, $\Delta p_\text{L} = p_\text{L} - p_0$, as a function of the maximum bubble radius, $R_\text{max}$, of coated bubbles with $R_0 \in \{1,2\} \, \mu \text{m}$ and $\kappa_\text{s} = 7.5 \times 10^{-9} \, \text{kg/s}$, simulated using the Gilmore equation. The hard-core radius $h$ is determined based on the properties of SF$_6$.}
\label{fig:Rmax-pL_compRdim}
\end{figure}

This scaling of the maximum pressure amplitude generated at the bubble wall, proportional to power $3/2$ of the dimensionless maximum bubble radius, may be understood from the energy balance of the bubble collapse. The transient energy of the radiated pressure wave when passing a virtual sphere surface at a distance $r$ from the bubble center is given by $E_\text{w} = 4\pi r^2 \int_{\tau_\text{w}}p^2\left(r,t\right) \, \text{d}t/ \left(\rho_\ell  c_\ell \right)$ \citep{Cole1948}, where $\tau_\text{w}$ is a measure of the wave passage time at $r$. With $\overline{p}^2\left(r\right)$ being the mean of $p^2\left(r,t\right)$ over $\tau_\text{w}$, the integral expression can be replaced by $\tau_\text{w} \overline{p}^2\left(r\right)$, and the wave energy becomes $E_\text{w}=4\pi r^2 \tau_\text{w} \overline{p}^2\left(r\right)/ \left(\rho_\ell c_\ell \right)$. \citet{Fortes-Patella2013} demonstrated that the shape of the dimensionless transient pressure signal $p\left(r,t\right)/p_\text{max}\left(r\right)$ as a function of the dimensionless time $t/\tau_\text{w}$ is invariant with respect to $R_0$ for the inertial collapse of vapour bubbles in a liquid at $p_\infty = 10^5 \, \text{Pa}$. This gives rise to the relation $\overline{p}^2\left(r\right) = \zeta p_\text{max}^2 \left(r\right)$, where $\zeta$ is the shape parameter of the transient pressure wave \citep{Fortes-Patella2013}. Furthermore, the pressure decay law for spherical shock waves \citep{Akulichev1971, Holzfuss2010} suggests that $p_\text{max}\left(r\right) \propto \max\{p_\text{L}\}$. For large excitation amplitudes, the initial potential energy content of the collapsing bubble is approximately given by $E_\text{pot,0} \approx 4\pi R_\text{max}^3 \Delta p_\text{a}/3$. Even though the pressure difference driving the bubble collapse is not constant throughout the bubble collapse for the periodically excited bubble, one may still assume that $E_\text{w} \propto E_\text{pot,0}$ and, hence, $p_\text{max}^2 \propto R_\text{max}^3$, which might explain why $\max\{\Delta p_\text{L}\} \propto (R_\text{max} / R_0-1)^{3/2}$ for large $R_\text{max} / R_0$. However, a more detailed analysis is required to clarify the exact origin of this scaling.

The maximum amplitude of the pressure generated in the liquid at the bubble wall is shown for different modelling assumptions in Figure \ref{fig:Rmax-pL_compProp}, simulated using the Gilmore equation.  The initial bubble radius $R_0$ clearly influences the pressure amplitude for $R_\text{max} \leq R_\text{rupt}$, where the bubble oscillations are in the linear regime, as well as the transition to the inertial regime, as evident in Figure \ref{fig:Rmax-pL_compProp}a. However, the initial size of the bubble becomes largely irrelevant for the relationship between dimensionless maximum bubble radius, $R_\text{max}/R_0-1$, and the amplitude of the generated pressure in the liquid, $\max\{\Delta p_\text{L}\}$, for large $R_\text{max} \gtrsim 6 \, \mu \text{m}$. Yet, since the maximum radius of a bubble at large excitation amplitudes is almost independent of the initial bubble radius \citep{Leighton1994}, the driving pressure difference is larger for smaller bubbles, which in turn generates emissions with larger pressure amplitudes by the bubble collapse. This can be observed in Figure \ref{fig:Rmax-pL_compRdim}, where the pressure amplitude generated by the bubble with $R_0 = 1 \, \mu \text{m}$ is approximately 3 times larger for a given maximum bubble radius than the pressure amplitude generated by the bubble with $R_0 = 2 \, \mu \text{m}$.

\begin{figure}[t]
  \centerline{\includegraphics[scale=1]{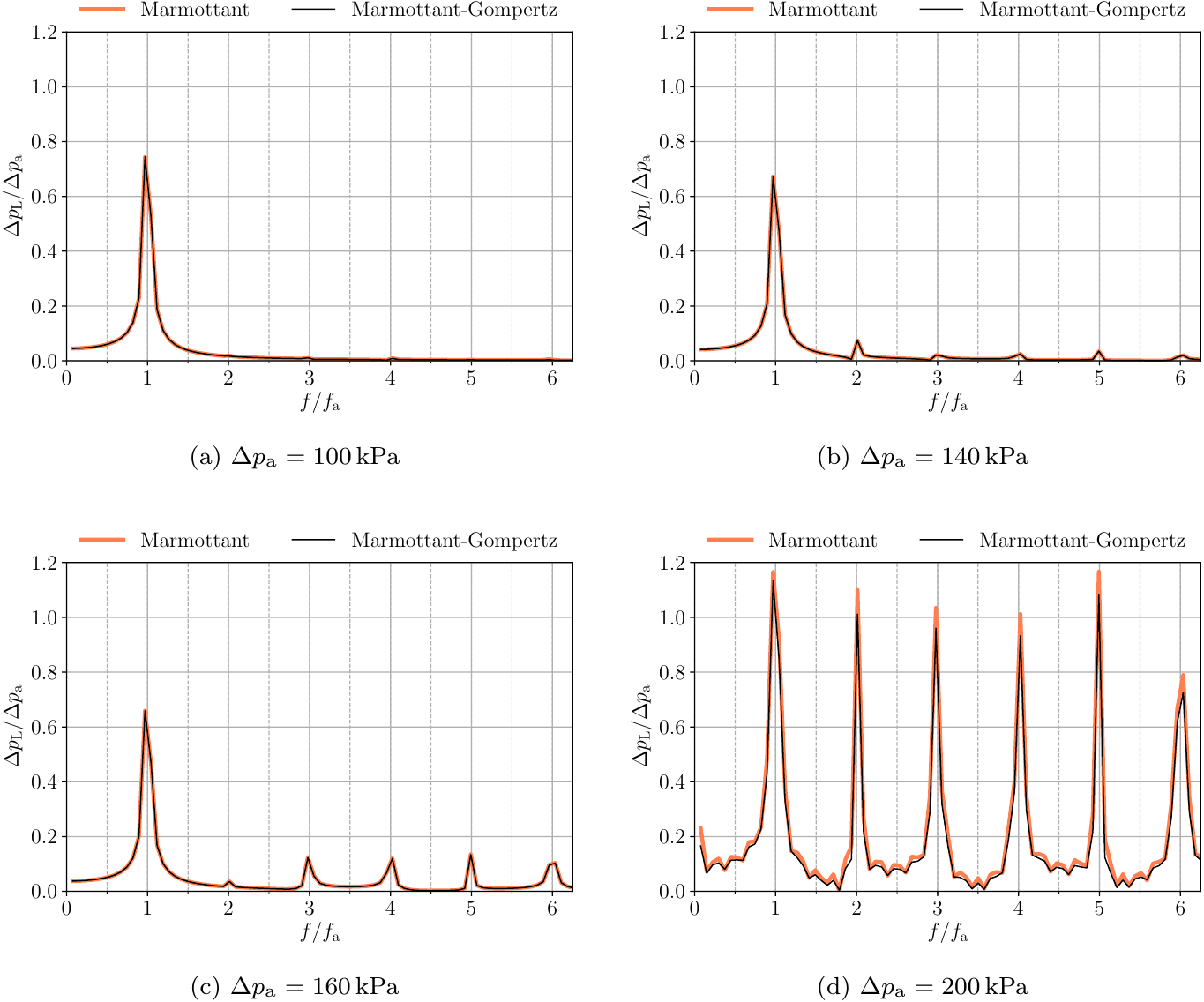}}
\caption{Dimensionless frequency spectrum of the liquid pressure at the bubble wall predicted by the Gilmore equation for the coated bubble ($\kappa_\text{s} = 7.5 \, \times 10^{-9} \, \text{kg/s}$) with $R_0 = 1.5 \, \mu\text{m}$ and $h = 310.4 \, \text{nm}$, excited with $f_\text{a} = 200 \, \text{kHz}$ and different excitation amplitudes $\Delta p_\text{a}$, using the Marmottant and the Marmottant-Gompertz lipid coating models.}
\label{fig:Marmottant-Gompertz_Pspec}
\end{figure}

The other modelling assumptions considered have evidently no significant influence on the maximum pressure generated by the bubble insonation for $R_\text{max} \leq R_\text{rupt}$. The surface dilatational viscosity affects the generated pressure predominantly during the transition from stable to inertial cavitation, see Figure \ref{fig:Rmax-pL_compProp}b, whereas the hard-core radius has a dominant influence on the generated pressure in the inertial regime, see Figure \ref{fig:Rmax-pL_compProp}c, if $R_\text{max} > 5 R_0$ for the bubble considered here. 
The continuous change of the surface tension coefficient introduced by the Marmottant-Gompertz model does not present any discernible differences in the generated liquid pressure amplitude compared to the Marmottant model, as seen in Figure \ref{fig:Rmax-pL_compProp}d.

\section*{Frequency spectrum of the acoustic emissions}

\begin{figure}[t]
  \centerline{\includegraphics[scale=1]{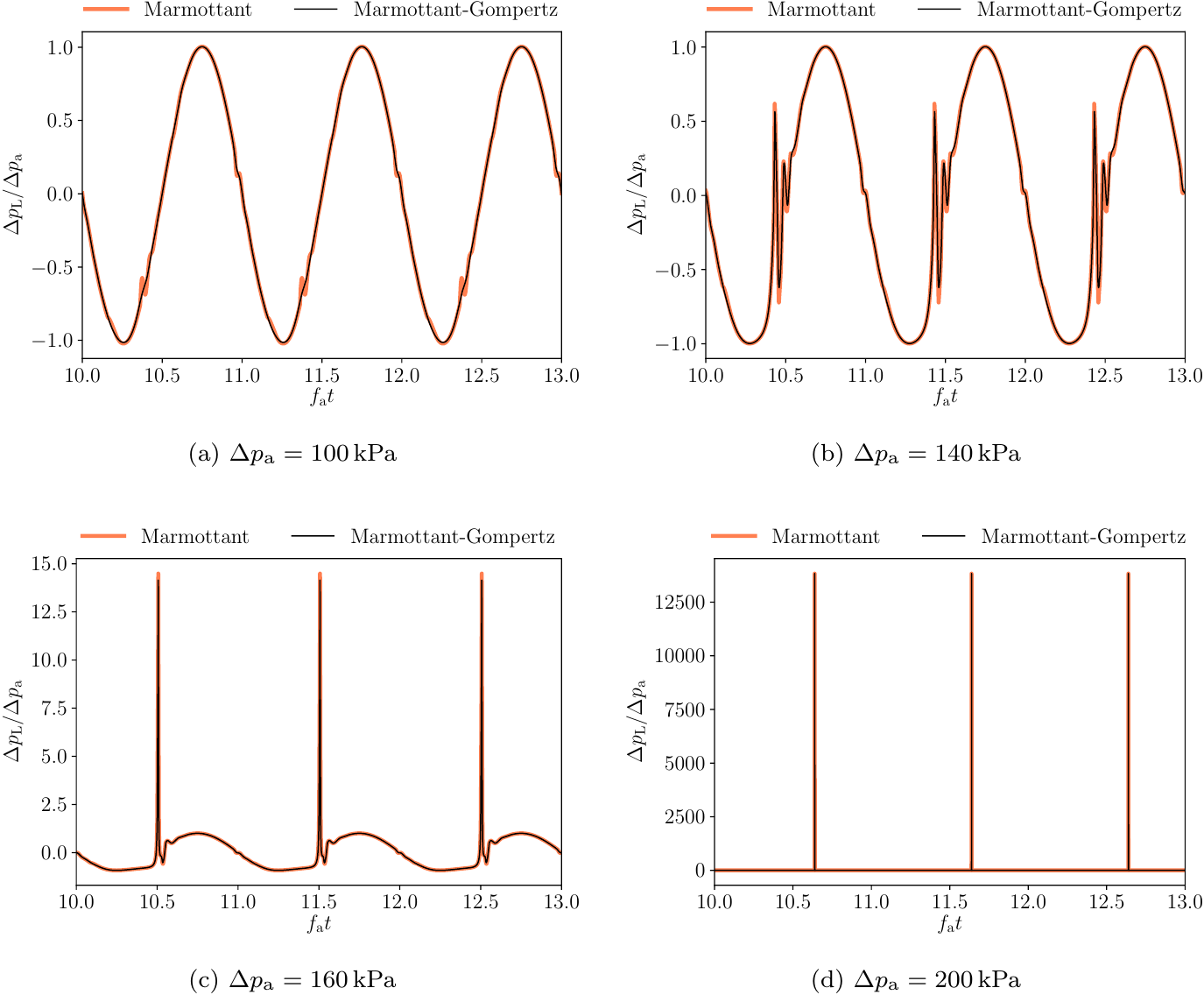}}
\caption{Dimensionless liquid pressure $\Delta p_\text{L}/\Delta p_\text{a}$ at the bubble wall as a function of time predicted by the Gilmore equation for the coated bubble ($\kappa_\text{s} = 7.5 \, \times 10^{-9} \, \text{kg/s}$) with $R_0 = 1.5 \, \mu\text{m}$ and $h = 310.4 \, \text{nm}$, excited with $f_\text{a} = 200 \, \text{kHz}$ and different excitation amplitudes $\Delta p_\text{a}$, using the Marmottant and the Marmottant-Gompertz lipid coating models.}
\label{fig:Marmottant-Gompertz_P}
\end{figure}

\begin{figure}[t]
  \centerline{\includegraphics[scale=1]{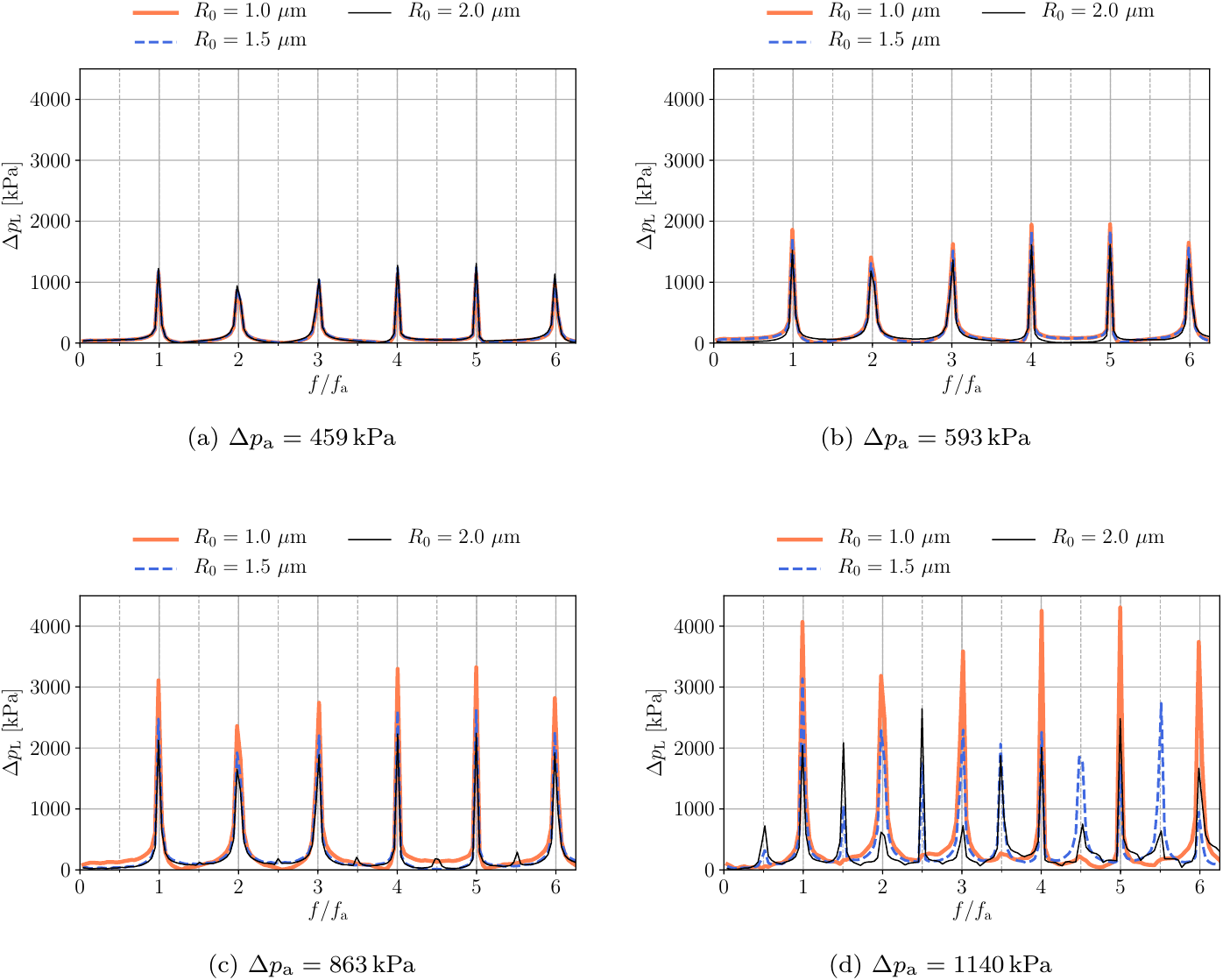}}
\caption{Frequency spectrum of the pressure wave generated in the liquid at the bubble wall predicted by the Rayleigh-Plesset equation for the coated bubble ($\kappa_\text{s} = 7.5 \, \times 10^{-9} \, \text{kg/s}$) with different initial bubble radii $R_0$, excited with $f_\text{a} = 692 \, \text{kHz}$ and different excitation amplitudes $\Delta p_\text{a}$.}
\label{fig:spectrum692RPAR}
\end{figure}

\begin{figure}[t]
  \centerline{\includegraphics[scale=1]{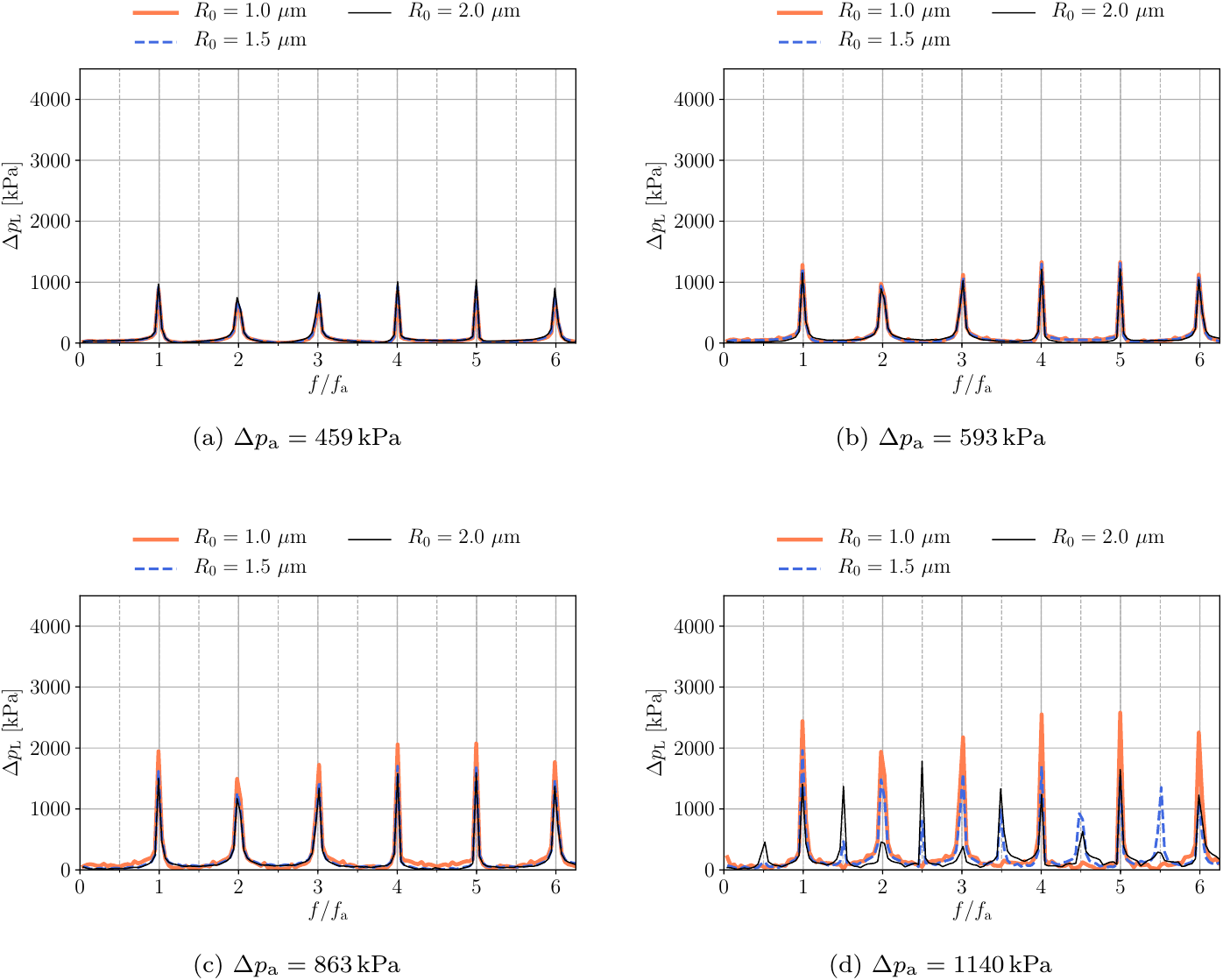}}
\caption{Frequency spectrum of the pressure wave generated in the liquid at the bubble wall predicted by the Gilmore equation for the coated bubble ($\kappa_\text{s} = 7.5 \, \times 10^{-9} \, \text{kg/s}$) with different initial bubble radii $R_0$ , excited with $f_\text{a} = 692 \, \text{kHz}$ and different excitation amplitudes $\Delta p_\text{a}$.}
\label{fig:spectrum692}
\end{figure}

Figure \ref{fig:Marmottant-Gompertz_Pspec} shows the frequency spectra of a coated bubble with $R_0 = 1.5 \, \mu \text{m}$ excited by the smallest excitation frequency considered, $f_\text{a} = 200 \, \text{kHz}$, at different excitation amplitudes $\Delta p_\text{a}$, using the Gilmore equation in conjunction with the Marmottant model or the Marmottant-Gompertz model. At $\Delta p_\text{a} = 100 \, \text{kPa}$, the bubble oscillates in a linear fashion at the excitation frequency, see Figure \ref{fig:Marmottant-Gompertz_Pspec}a. The evolution of the pressure at the bubble wall is predominantly sinusoidal, as seen in Figure \ref{fig:Marmottant-Gompertz_P}a, which shows the normalised liquid pressure at the bubble wall as a function of time. A small difference in the pressure evolution can be observed between the Marmottant model, which exhibits pressure discontinuities when entering and leaving the elastic regime ($R_\text{buck} <R< R_\text{rupt}$), and the Marmottant-Gompertz model, which does not exhibit these discontinuities. When the excitation amplitude is increased to $\Delta p_\text{a} = 140 \, \text{kPa}$, the bubble is still in the linear regime as seen in Figure \ref{fig:Marmottant-Gompertz_P}b, but higher harmonics ($nf_\text{a}$ with $n = 2,3,...$) start to emerge in the frequency spectrum shown in Figure \ref{fig:Marmottant-Gompertz_Pspec}b. For further increasing excitation pressure amplitudes the peak pressure at the bubble wall begins to increase nonlinearly, and for $\Delta p_\text{a} = 200 \, \text{kPa}$ the bubble response is dominated by inertia, as evident by the dominant higher harmonics in the frequency spectrum in Figure \ref{fig:Marmottant-Gompertz_Pspec}d and the sharp pressure peaks in Figure \ref{fig:Marmottant-Gompertz_P}d. This sudden increase of the amplitude of higher harmonics at the onset of inertial cavitation agrees with experimental measurements of \citet{Ilovitsh2018}, who observed a nonlinear increase of the amplitude of the second harmonic at the onset of inertial cavitation. Overall, the discontinuous definition of the surface tension coefficient by the Marmottant model has no significant influence on the frequency spectra shown in Figure \ref{fig:Marmottant-Gompertz_Pspec}.

\begin{figure}[t]
  \centerline{\includegraphics[scale=1]{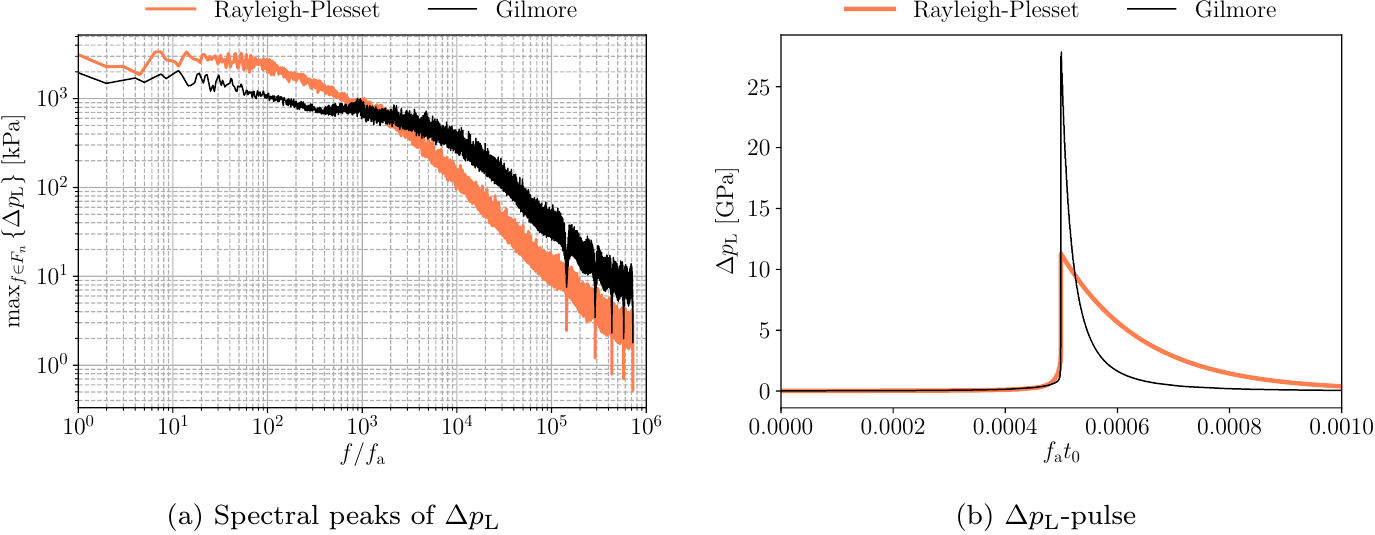}}
\caption{Evolution of (a) the spectral peak values of $\Delta p_\text{L}$ and (b) the pressure pulse associated with the largest value of $\Delta p_\text{L}$ for a coated bubble ($\kappa_\text{s} = 7.5 \times 10^{-9} \, \text{kg/s}$) with $R_0 = 1.5 \, \mu\text{m}$ and $h = 310.4 \, \text{nm}$, excited at $f_\text{a} = 692 \, \text{kHz}$ and $\Delta p_\text{a} = 1140 \, \text{kPa}$, solved with the Rayleigh-Plesset equation and the Gilmore equation. The spectral peaks are given by the maximum values of $\Delta p_\text{L}$ in subsequent frequency windows $F_n=\left[f_n, f_n+1.3f_\text{a}\right]$ in the frequency domain. The pressure pulses are shifted relative to each other such that the peak values collapse on $f_\text{a}t_0= 5 \times 10^{-4}$ and the window length corresponds to the dimensionless frequency $f/f_\text{a}=10^3$.}
\label{fig:spectralPeaks_RPvsGilmore}
\end{figure}

\begin{figure}[t]
  \centerline{\includegraphics[scale=1]{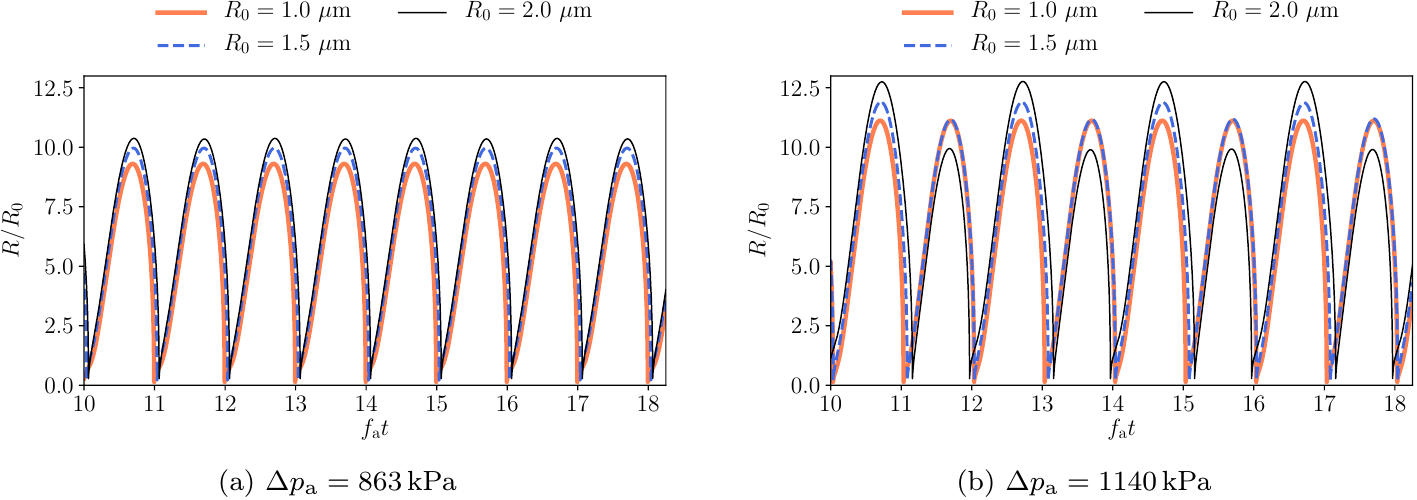}}
\caption{Dimensionless bubble radius $R/R_0$ as a function of the dimensionless time $f_\text{a} t$ for a coated bubble ($\kappa_\text{s} = 7.5 \times 10^{-9} \, \text{kg/s}$) with different initial radii $R_0 \in \{1.0,1.5,2.0\} \, \mu\text{m}$ excited at $f_\text{a} = 692 \, \text{kHz}$ and different excitation pressure amplitudes $\Delta p_\text{a} \in \{863 ,1140\} \, \text{kPa}$, predicted by the Gilmore equation in conjunction with the Marmottant model.}
\label{fig:Rvt}
\end{figure}

\begin{figure}[t]
\centerline{\includegraphics[scale=1]{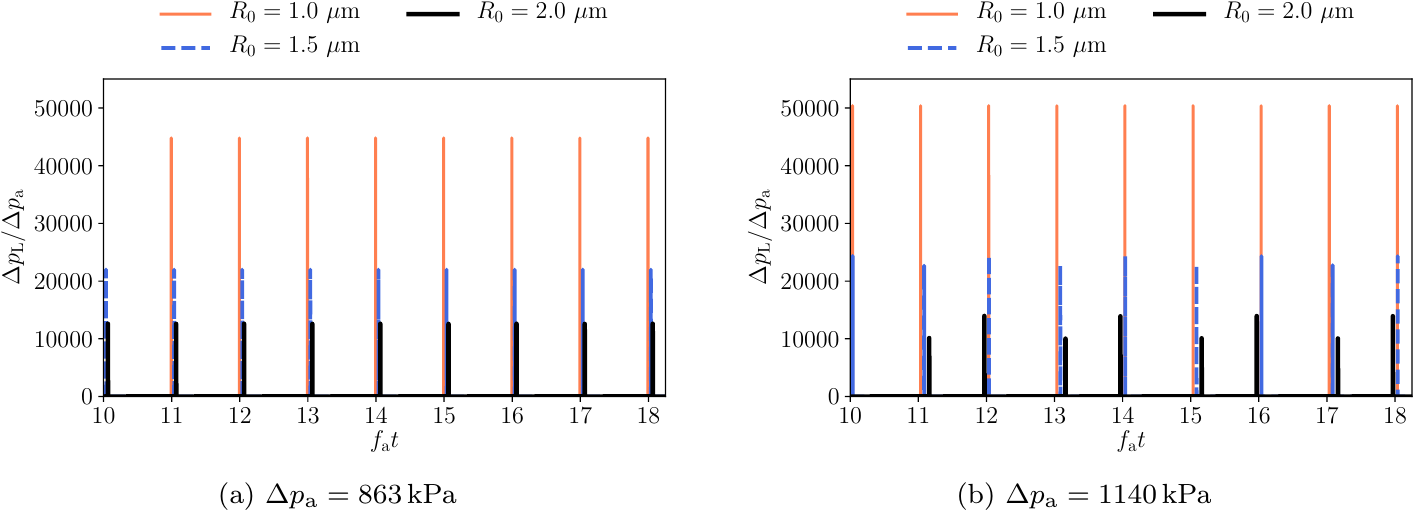}}
\caption{Dimensionless liquid pressure $\Delta p_\text{L}/\Delta p_\text{a}$ at the bubble wall as a function of the dimensionless time $f_\text{a} t$ for a coated bubble ($\kappa_\text{s} = 7.5 \times 10^{-9} \, \text{kg/s}$) with different initial radii $R_0 \in \{1.0,1.5,2.0\} \, \mu\text{m}$ excited at $f_\text{a} = 692 \, \text{kHz}$ and different excitation pressure amplitudes $\Delta p_\text{a} \in \{863 ,1140\} \, \text{kPa}$, predicted by the Gilmore equation in conjunction with the Marmottant model.}
\label{fig:pvt}
\end{figure}

\begin{figure}[t]
  \centerline{\includegraphics[scale=1]{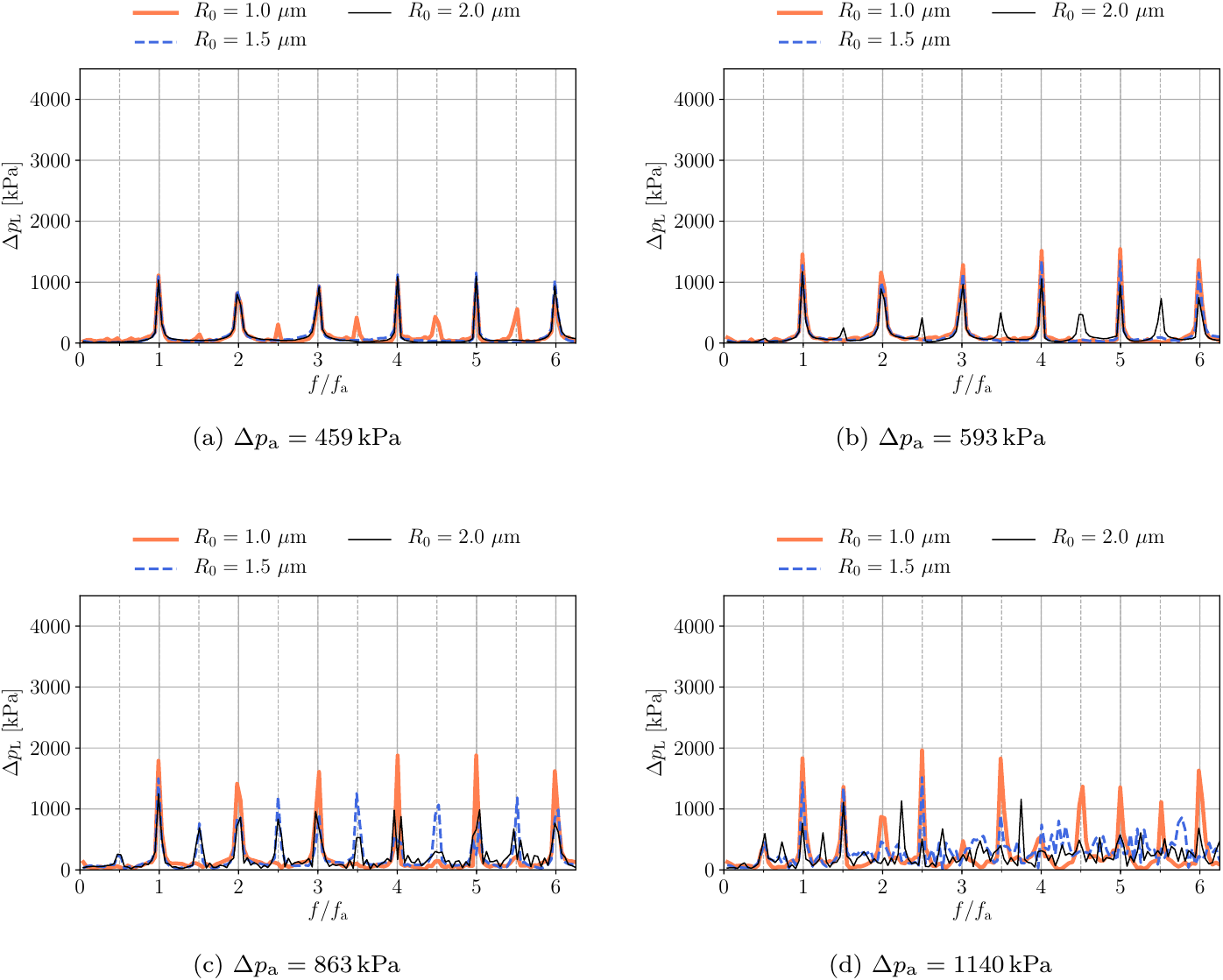}} 
\caption{Frequency spectrum of the pressure wave generated in the liquid at the bubble wall predicted by the Gilmore equation for the clean bubble with different initial bubble radii $R_0$ excited with $f_\text{a} = 692 \, \text{kHz}$ and different excitation amplitudes $\Delta p_\text{a}$.}
\label{fig:spectrum692clean}
\end{figure}

\citet{Song2019} recently reported {\em in vitro} hydrophone measurements of the pressure signal, including the frequency spectrum, emitted by SonoVue bubbles excited with a frequency of $f_\text{a} = 692 \, \text{kHz}$ and four different pressure amplitudes, $\Delta p_\text{a} \in \{459, 593, 863, 1140\} \, \text{kPa}$. \citet{Song2019} stated that cavitation activity in their experiment was initiated by bubbles that were unresolved by the high-speed imaging, which in turn was reported to have a resolution of $4.2 \, \mu \text{m}$ per pixel. We, therefore, assume that the spectra reported by \citet{Song2019} correspond to bubbles with $R_0 < 2.1 \, \mu \text{m}$. Interestingly, the hydro\-phone measurements recorded subharmonic and ultraharmonic emissions only for the largest excitation pressure amplitude, $\Delta p_\text{a} = 1140 \, \text{kPa}$, but not for the three smaller pressure amplitudes. Figures \ref{fig:spectrum692RPAR} and \ref{fig:spectrum692} show the frequency spectra of the four cases excited with $f_\text{a} = 692 \, \text{kHz}$ computed by the Rayleigh-Plesset equation and the Gilmore equation, respectively, for the coated bubble with different initial radii $R_0$. 
Both the Rayleigh-Plesset equation and the Gilmore equation predict the lack of sub- and ultraharmonic emissions for $\Delta p_\text{a} \in \{459,593,863\} \, \text{kPa}$ and, especially for $R_0 = 1.5 \, \mu \text{m}$ and $R_0 = 2.0 \, \mu \text{m}$, the occurrence of sub- and ultraharmonic emissions for $\Delta p_\text{a} = 1140 \, \text{kPa}$, in agreement with the measurements of \citet{Song2019}.
Despite the large Mach numbers, $M_\ell \gtrsim 1$, occurring in all four cases, the Rayleigh-Plesset equation predicts the main qualitative features of the spectra, in good agreement with the Gilmore equation.

However, the spectral peaks predicted by the Rayleigh-Plesset equation have considerably different amplitudes compared to the spectral peaks predicted by the Gilmore equation. According to Figures \ref{fig:spectrum692RPAR} and \ref{fig:spectrum692}, the Rayleigh-Plesset equation predicts significantly larger spectral peaks than the Gilmore equation for the first six harmonics. In contrast, Figure \ref{fig:liqPressureLargeAmp} indicates that the Gilmore equation predicts larger maximum values of $\Delta p_\text{L}$ for the conditions at hand. This behaviour is explained by Figure \ref{fig:spectralPeaks_RPvsGilmore}, which shows the decay of the spectral peaks of $\Delta p_\text{L}$ and a representative pressure pulse at the bubble wall for both the Rayleigh-Plesset equation and the Gilmore equation. The spectral peak curve in Figure \ref{fig:spectralPeaks_RPvsGilmore}a represents the peak values of $\Delta p_\text{L}$ in subsequent frequency windows of length $1.3f_\text{a}$, such that each window can be expected to contain at least one spectral peak. It can be seen that while the Rayleigh-Plesset equation produces significantly larger spectral peaks for $f/f_\text{a} \lesssim 800$, the Gilmore equation produces larger spectral peaks for $f/f_\text{a} \gtrsim 1200$ at a lower magnitude, however over a significantly larger frequency range. The high-frequency spectral content involves harmonics with periods that are in the order or smaller than the pressure pulse duration, and therefore forms a significant contribution to the shape and the amplitude of the pressure pulse. This can be seen in Figure \ref{fig:spectralPeaks_RPvsGilmore}b, where the Gilmore equation predicts a pressure pulse of shorter duration but higher amplitude than the Rayleigh-Plesset equation. For reference, the dimensionless time window length in Figure \ref{fig:spectralPeaks_RPvsGilmore}b is chosen to be $10^{-3}$, which corresponds to the dimensionless frequency $f/f_\text{a}=10^3$, which again marks the approximate intersection of the spectral peak curves in Figure \ref{fig:spectralPeaks_RPvsGilmore}a.

The subharmonic and ultraharmonic contributions at $\Delta p_\text{a} = 1140 \, \text{kPa}$ were attributed by \citet{Song2019} to a period-doubled collapse of this bubble, which can also be observed in Figure \ref{fig:Rvt}b for the bubbles with $R_0 = 1.5 \, \mu \text{m}$ and $R_0 = 2.0 \, \mu \text{m}$ excited at $\Delta p_\text{a} = 1140 \, \text{kPa}$. No period doubling is observed in Figure \ref{fig:Rvt}a for $\Delta p_\text{a} = 863 \, \text{kPa}$. However, contrary to \citet{Song2019}, who reported the emission of shock waves at alternate compression phases, a sharp pressure peak reminiscent of a shock wave is emitted, albeit with slightly different amplitude, during every compression phase, as seen in Figure \ref{fig:pvt}.

If the bubble is considered to be clean, {\em i.e.}~if the lipid coating is not accounted for, for which the spectra are shown in Figure \ref{fig:spectrum692clean}, the onset of sub- and ultraharmonic contributions occurs at a lower excitation pressure amplitude and are already clearly visible for the bubbles with $R_0 = 1.5 \, \mu \text{m}$ and $R_0 = 2.0 \, \mu \text{m}$ excited at  $\Delta p_\text{a} = 863 \, \text{kPa}$, with significantly more sub- and ultraharmonics present in the spectrum of the liquid pressure generated by the bubble excited with $\Delta p_\text{a} = 1140 \, \text{kPa}$. Hence, the lipid coating has a critical influence on the frequency spectrum of the generated pressure wave.

\section*{Discussion}

\subsection*{Validity and limitations of the modelling assumptions}

The main limitations associated with the employed modelling approach pertain to the sphericity of the bubble, the uncertainty of the properties and condition of the lipid monolayer, and neglecting thermal effects.

The single-bubble Rayleigh-Plesset-type models used in this study assume the bubble to be spherical throughout the simulation. Fragmentation of the bubble for sufficiently large excitation amplitudes \citep{Chomas2001} is not modelled and, as a consequence, changes in acoustic emissions resulting from a reducing bubble size cannot be predicted. The single-bubble models also do not account for the interaction of the modelled microbubble with neighbouring bubbles or with other objects, such as the vasculature. It is well established that such interactions can lead to nonspherical bubble oscillations, which promote fragmentation \citep{Chomas2001}, and asymmetric bubble collapse \citep{Cleve2019}, which can cause cell lysis \citep{Chen2003c} and sonoporation \citep{Prentice2005, Ohl2006}. Nonspherical bubble oscillations may also form for a single isolated bubble as a result of shape instabilities \citep{Dollet2008,Lajoinie2018}. Furthermore, changes of the properties of the lipid monolayer caused, for instance, by the duration of the insonation \citep{Qin2009}, a locally increased lipid concentration \citep{Kooiman2017} or the shedding of lipids from the monolayer \citep{Borden2002, Lajoinie2018}, are not accounted for.

The extreme pressures, even predicted to exceed $100 \, \text{GPa}$ for many cases, raise questions regarding the validity of the equations of state used to describe the gas and the liquid. For the gas, our results demonstrate an important influence of the hard-core radius, which limits the compressibility of the gas and, consequently, the pressure inside the bubble. 
The Tait equation of state used in the Gilmore equation \citep{Gilmore1952} to describe the thermodynamic properties of the liquid is known for yielding unrealistic heat capacities, which influences the predicted enthalpy at the bubble wall and the velocity of the bubble wall \citep{Denner2021}.
Based on equilibrium thermodynamics, the fluid may also reach a supercritical state at such high pressures or, in the case of water, may even solidify, which would require entirely different physical models and equations of state. However, given the very short time spans of less than $500 \, \mathrm{ps}$ over which these extreme pressure amplitudes typically occur, as observed in Figure \ref{fig:spectralPeaks_RPvsGilmore}b where the shown time interval corresponds to $877 \, \text{ps}$, we conjecture that the thermodynamic system is unable to establish a thermodynamic equilibrium. Studies on the modelling of sonoluminescence, where similarly large pressure values and bubble wall velocities are observed, corroborate the overall applicability of the modelling assumptions used in this study to predict the bubble dynamics \citep{Vignoli2013, Nazari-Mahroo2018}.

A strong bubble collapse with large pressure amplitudes also leads to extreme gas temperatures, with maximum values of  $\mathcal{O}(10^4) \, \text{K}$. The Peclet number of the gas, $\text{Pe}_\text{g} = 2 \pi f_\text{a} R_0^2/\alpha_\text{g}$, where $\alpha_\text{g}$ is the thermal diffusivity of the gas, is $\text{Pe}_\text{g}>1$ for the bubble with $R_0 = 1.5 \, \mu \text{m}$ across the considered excitation regime. Thus, the adiabatic assumption for the gas employed in this study is reasonable \citep{Bergamasco2017}. However, the Peclet number is $\mathcal{O}(1)$ for the lower frequencies considered in this study, and even $\text{Pe}_\text{g} < 1$ for $R_0 = 1 \, \mu \text{m}$, for which heat conduction between the gas and the liquid should, thus, be expected to have an influence on the bubble behaviour. Quantifying this influence requires dedicated studies.

The Rayleigh-Plesset equation has been found to predict Mach numbers of the bubble wall of $M \simeq 1$ and above, placing it outside its region of formal validity, for large parts of the considered acoustic excitation regime, whereas the Gilmore equation is formally valid throughout the considered acoustic excitation regime. Our results identify a Mach number of the bubble wall of $M \simeq 0.1$ as the point where the predictions of the Rayleigh-Plesset equation and the Gilmore equation start to deviate considerably from each other. 

Despite the numerous simplifying assumptions underpinning the conducted simulations, an encouraging agreement is observed for key quantities. 
For the particular case of a bubble with $R_0 = 1.5 \, \mu \text{m}$, the onset of inertial cavitation (or, likewise, the departure from quasi-linear bubble behaviour) is found to occur for a pressure amplitude of approximately $150 \, \text{kPa}$, which is in very good agreement with {\em in vitro} experiments of \citet{Ilovitsh2018} using a single lipid-coated bubble of the same size. This agreement is further substantiated by the fact that, apart from the initial bubble size, none of the considered modelling assumptions has a significant impact on the onset of inertial cavitation; small differences in lipid-coating properties or type of gas core do not affect this comparison. The excitation pressure amplitude required for the onset of sub- and ultraharmonic acoustic emissions is in very good agreement with {\em in vitro} measurements of SonoVue bubbles by \citet{Song2019}. Notably, at the onset of sub- and ultraharmonic emissions for the considered bubble sizes and excitation frequencies, the bubble behaviour is nonlinear. Hence, the single-bubble models employed in this study are able to predict key quantities in the (quasi-)linear regime (onset of inertial cavitation) and in the nonlinear regime (onset of sub- and ultraharmonic emissions).

\subsection*{Marmottant-Gompertz model}

\begin{figure}[t]
  \centerline{\includegraphics[scale=1]{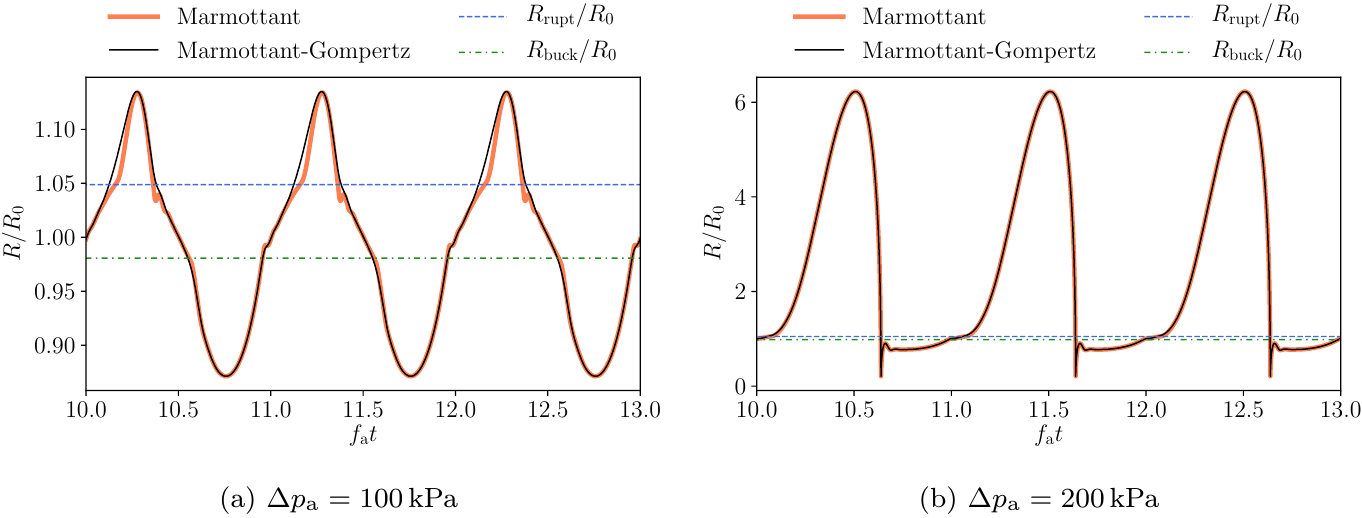}}
  \caption{Dimensionless bubble radius $R/R_0$ as a function of the dimensionless time $f_\text{a} t$ for a coated bubble ($\kappa_\text{s} = 7.5 \times 10^{-9} \, \text{kg/s}$) with  $R_0 = 1.5 \, \mu\text{m}$ excited at $f_\text{a} = 200 \, \text{kHz}$ and different excitation pressure amplitudes $\Delta p_\text{a} \in \{100 ,200\} \, \text{kPa}$, predicted by the Gilmore equation in conjunction with the Marmottant model and the Marmottant-Gompertz model. The buckling radius, $R_\text{buck}$, and the rupture radius, $R_\text{rupt}$, of the lipid monolayer are shown as a reference.}
  \label{fig:Marmottant-Gompertz_R}
  \end{figure}

In order to assess the influence of the two discontinuities of the surface tension coefficient embedded in the original lipid-coating model of \citet{Marmottant2005}, we approximate the change in surface tension coefficient in function of the bubble radius by a Gompertz function, see Eq.~(\ref{eq:sigma_gompertz}), using the same input parameters as for the Marmottant model. The surface tension coefficient given by this Marmottant-Gompertz model is continuous and differentiable, as seen in Figure \ref{fig:gompertz} for representative lipid monolayer properties.
The pressure amplitude at the bubble wall, shown in Figure \ref{fig:Marmottant-Gompertz_P}, exhibits a smoother pressure evolution for $\Delta p_\text{a} = 100 \, \text{kPa}$ with the Marmottant-Gompertz model than with the Marmottant model, yet no significant differences between the Marmottant model and the Marmottant-Gompertz model are observed for the acoustic emissions of the considered cases.
Figure \ref{fig:Marmottant-Gompertz_R} shows the evolution of the dimensionless radius, $R/R_0$, for a coated bubble with $R_0 = 1.5 \, \mu \text{m}$, excited with $f_\text{a} = 200 \, \text{kHz}$ and $\Delta p_\text{a} \in \{100,200\} \, \text{kPa}$. Exciting the bubble with $\Delta p_\text{a} = 100 \, \text{kPa}$, the bubble spends a considerable share of each excitation period in the elastic regime ($R_\text{buck} < R < R_\text{rupt}$). While the discontinuous change in surface dilatational modulus manifests as discontinuities in the evolution of the bubble radius with the Marmottant model, the bubble radius transitions smoothly in and out of the elastic regime when the Marmottant-Gompertz model is used. However, the overall dynamics of the bubble remain unaffected. Exciting the bubble at a larger amplitude of $\Delta p_\text{a} =200 \, \text{kPa}$, the results obtained with the two models become virtually indistinguishable. The Marmottant-Gompertz model can, therefore, serve as a simple extension of the original Marmottant model that provides a continuous definition of the surface tension coefficient and, consequently, the surface dilatational modulus of the lipid monolayer coating using the same input parameters, without affecting the overall behaviour of the bubble.

\subsection*{Linear bubble response}

The presented results suggest a linear relationship between the excitation pressure amplitude and the maximum pressure generated in the liquid. Especially for small excitation frequencies and small initial bubble radii, this linear regime is delineated sharply from the onset of nonlinear behaviour, whereas larger excitation frequencies or larger initial bubble radii promote a smoother transition from linear to nonlinear behaviour.
The Blake pressure as given in Eq.~(\ref{eq:pblake}) is found to be a reliable estimate for the onset of nonlinear bubble behaviour, even though Eq.~(\ref{eq:pblake}) does not account for the lipid coating.

\begin{figure}[t]
  \centerline{\includegraphics[scale=1]{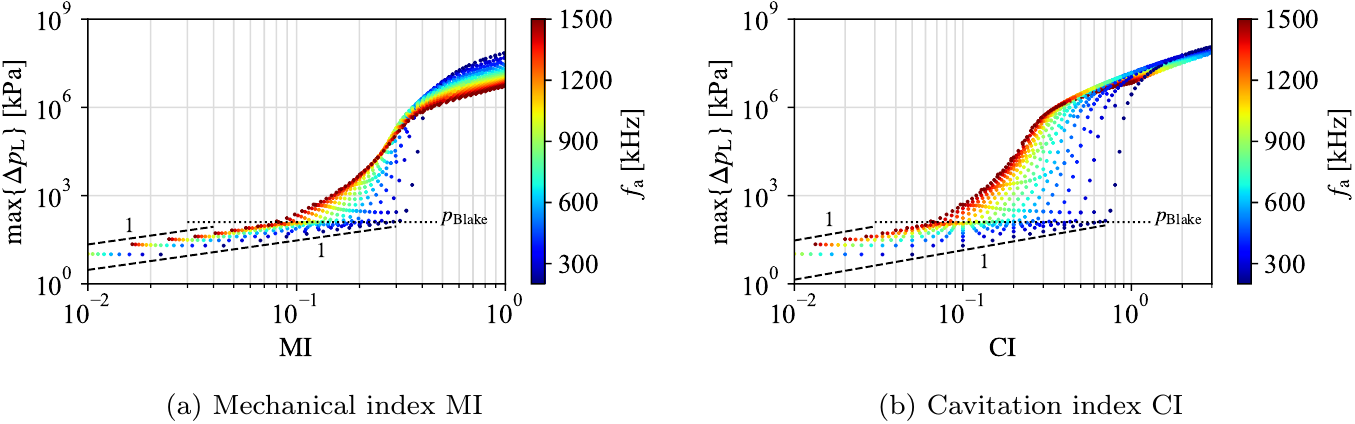}}
\caption{The maximum liquid pressure at the bubble wall, $\Delta p_\text{L} = p_\text{L} - p_0$, of a coated bubble with $R_0= 1.5  \, \mu \text{m}$, $\kappa_\text{s} = 7.5 \times 10^{-9} \, \text{kg/s}$ and $h = 310.4 \, \text{nm}$ predicted by the Gilmore equation, as a function of the mechanical index $\text{MI}$ and the cavitation index $\text{CI}$. The colour of the data points represents the excitation frequency $f_\text{a}$. The scaling exponent of $1$ represents a linear relationship between the maximum pressure amplitude in the liquid at the bubble wall and the corresponding index, and $\Delta p_\text{L} = p_\text{Blake}$ is shown as a reference.}
\label{fig:transitionMICI}
\end{figure}

The mechanical index \citep{Apfel1991}, $\text{MI} = \text{PNP}/\sqrt{f_\text{a}^\star}$, is frequently used to predict and categorise bioeffects caused by cavitation, where $\text{PNP}$ is the peak negative pressure of the acoustic excitation in MPa and $f_\text{a}^\star$ is the excitation frequency $f_\text{a}$ in MHz. In addition, the cavitation index \citep{Bader2013}, $\text{CI} = \text{PNP}/f_\text{a}^\star$, was proposed with the aim of predicting subharmonic emissions and rupture of the bubble coating. The maximum pressure amplitude generated in the liquid at the bubble wall for a coated bubble with $R_0= 1.5  \, \mu \text{m}$ is shown in Figure \ref{fig:transitionMICI}, as a function of both the mechanical index and the cavitation index. The maximum pressure amplitude in the liquid, $\max\{\Delta p_\text{L}\}$, depends linearly on both indices for sufficiently small excitation pressure amplitudes $\Delta p_\text{a}$ (expected, since $\text{MI} \propto \Delta p_\text{a}$ and $\text{CI} \propto \Delta p_\text{a}$) and this linear regime is bounded consistently by the Blake pressure, $p_\text{Blake}$, as indicated in Figure \ref{fig:transitionMICI}. 

The widely used approximate threshold for inertial cavitation and the onset of adverse bioeffects is $\text{MI} =0.4$ \citep{Qin2009, Bader2013}, which is typically considered to be the upper limit for contrast-enhanced ultrasound imaging \citep{Tang2011, terHaar2021}. For the considered lipid-coated bubble shown in Figure \ref{fig:transitionMICI}, $\text{MI} = 0.4$ marks the end of the transition from stable to inertial cavitation. Other studies postulated a (nondestructive) weakly nonlinear response of coated microbubbles for $\text{MI}<0.1$ and a (destructive) strongly nonlinear response for $\text{MI} > 0.3$ \citep{Correas2001, Sennoga2017}; Figure \ref{fig:transitionMICI}a exhibits a comparable delineation of these regimes.

The linear regime identified in the presented results is particularly interesting for a concurrent FUS treatment and quantitative imaging of the region of interest, especially in conjunction with low frequency excitation that provides low attenuation of the ultrasound beam and, in cerebral applications, good transcranial transmission. 
The linear response of the bubble is particularly amenable to a closed-loop feedback control; a small change in excitation pressure amplitude (or mechanical index) corresponds to a linear change in emitted pressure amplitude that can be easily identified. Furthermore, linear emissions allow a superposition of the emitted acoustic waves, which is preferred for quantitative imaging techniques \citep{Tang2011, Krix2021}. 
To this end, our results point to low excitation frequencies in the range $200 \, \text{kHz} \leq f_\text{a} \leq 400 \, \text{kHz}$, and small bubbles, $R_0 \leq 1.5 \, \mu \text{m}$, as particularly suitable, because this results in a clear delineation of the linear and nonlinear regimes, with an extended pressure range of the linear regime for smaller bubbles. However, we should caution that this is a conjecture, since our results only consider a single bubble, without accounting for any interactions with other bubbles or, for instance, the vasculature. Moreover, the influence of tissue, the echo of which is known to contain harmonic multiples of the excitation (center) frequency \citep{Qin2009}, and material boundaries ({\em e.g.}~tissue-bone interfaces) on this linear regime is presently not clear.

From a computational viewpoint, this linear regime is well-suited for numerical analysis, since the bubble is unlikely to fragment \citep{Chomas2001} and extreme pressure amplitudes do not occur. Yet, the linear regime may be sensitive to the properties of the lipid monolayer, as observed in Figure \ref{fig:Marmottant-Gompertz_R}a, although the presented results do not exhibit any appreciable effect on the studied quantities.

\subsection*{Nonlinear bubble response}

Contrary to the linear regime, in the nonlinear regime the simplifying assumptions underpinning the employed numerical approach may have an influence on the presented results. In particular the large pressure amplitudes of $\mathcal{O}(10^{11}) \, \text{Pa}$ warrant further examination using more comprehensive numerical methods. Nevertheless, the maximum liquid pressure generated at the bubble wall can be estimated based on the maximum radius of the bubble, a quantity which can be determined optically with relatively high precision in experiments. Our analysis suggests that the 3/2-scaling between the maximum pressure generated at the bubble wall and the dimensionless bubble radius is associated with the energy balance of the bubble collapse. Since the bubble collapse in this regime is dominated by inertia, we do not expect the inclusion of additional physical mechanisms, such as heat transfer, to fundamentally change this relationship.
The onset of subharmonic and ultraharmonic frequencies of the liquid pressure signals predicted by both the Rayleigh-Plesset equation and the Gilmore equation are in good agreement with the experiments of \citet{Song2019}, which demonstrates that either of the considered primary equations in conjunction with the model by \citet{Marmottant2005} is potentially able to predict the frequency spectrum generated by lipid-coated microbubbles excited at subresonance frequencies. This is especially interesting with regards to the Rayleigh-Plesset equation, since it is far outside its formal range of validity for these cases. Accounting for the lipid coating has a critical influence on the amplitude and the frequency spectrum of the pressure waves generated in the liquid. 

Although the onset of subharmonic and ultraharmonic emissions agrees well between the numerical predictions in Figures \ref{fig:spectrum692RPAR} and \ref{fig:spectrum692} and the experiments of \citet{Song2019}, the amplitude of the pressure signal is more difficult to assess quantitatively. 
For instance, the pressure wave generated for $f_\text{a}=692 \, \text{kHz}$ and $\Delta p_\text{a} = 1140 \, \text{kPa}$ is strongly nonlinear, as evident by the broadband frequency spectrum shown in Figure \ref{fig:spectrum692}, and can reasonably be expected to form shock waves. Thus, as this pressure wave propagates spherically outward, its amplitude  does not reduce proportional to the inverse of the radial distance to the bubble with $1/r$, as dictated by the geometric divergence of a linear wave, such as a sound wave. Rather, this nonlinear pressure wave diminishes with $1/r^n$, where $1 \leq n \leq 2$ \citep{Akulichev1971, Holzfuss2010}, due to geometric divergence and the dissipation associated with the shock wave. 
For the four considered cases shown in Figures \ref{fig:spectrum692RPAR} and \ref{fig:spectrum692} with excitation frequency $f_\text{a} = 692 \, \text{kHz}$ adopted from the experiments of \citet{Song2019}, the Rayleigh-Plesset equation and the Gilmore equation predict a maximum liquid pressure that is roughly $1.8 \times 10^4$ to $1.8 \times 10^5$ times larger than the pressure measured by \citet{Song2019}. Considering the experimental measurement position is situated at a distance of approximately $350 \, \mu \text{m}$ from the cavitation site, as reported by \citet{Song2019}, and a minimum radius of the bubble of $R_\text{min} \simeq h$, this leads to $1.4 \lesssim n \lesssim 1.7$ for the relationship $p \propto 1/r^n$, which is within the theoretical limits. However, a detailed assessment of the emission spectra would have to include the propagation of the pressure wave in the liquid as well as any structures or objects between the cavitation site and the measurement site, {\em e.g.} the polycarbonate capillary in which the microbubbles were situated in the experiment.

\subsection*{Enabling a quantitative validation}

To enable a more detailed quantitative analysis of the predictive accuracy of Rayleigh-Plesset-type models in terms of acoustic emissions, more comprehensive computational models as well as more and carefully controlled {\em in vitro} studies are required. 

For such a comparison between experiments and simulations, our results point to the onset and amplitude of subharmonic and ultraharmonic frequencies of the pressure wave generated in the liquid as particularly attractive benchmarks for two reasons: (i) the onset of subharmonic and ultraharmonic frequencies are evidently sensitive to the modelling assumptions, and (ii) the spectral content of relatively low frequencies, particularly subharmonic, fundamental harmonic and the first ultraharmonic frequencies, decay slowly and are robust to interferences potentially arising in the experiments. Especially at large excitation amplitudes, above $\Delta p_\text{a} = 1140 \, \text{kPa}$ as considered by \citet{Song2019}, additional experiments likely hold new insights. 

A proper quantitative comparison further requires to simulate the full temperature distribution as well as the pressure distribution in the liquid coupled with the bubble dynamics. For the large pressure amplitudes predicted to be generated at small excitation frequencies in conjunction with large excitation pressure amplitudes, a suitable thermoacoustic model should also account for the thermodynamic coupling between the emitted pressure waves and the energy conservation, which in turn requires a more appropriate equation of state for the liquid than used in the classical Gilmore equation \citep{Radulescu2020, Denner2021}. Moreover, material boundaries between the cavitation site and the measurement location, {\em e.g.}~the wall of a capillary tube, and the presence of other bubbles may also need to be considered.

\subsection*{Real-time predictions}

Single-bubble models may not only be helpful in understanding the fundamentals of acoustically-driven bubble behaviour and estimating quantities associated with this behaviour, but also soon be able to serve as predictive tools for {\em in situ} treatment control. For instance, solving the Rayleigh-Plesset equation (\ref{eq:modRP}) for the SonoVue bubble with $R_0 = 1.5 \, \mu \text{m}$, $h = 310.4 \, \text{nm}$  and $\kappa_\text{s} = 7.5 \times 10^{-9}  \, \text{kg/s}$ considered in this study, excited with $f_\text{a} = 200 \, \text{kHz}$ and $\Delta p_\text{a} = 150 \, \text{kPa}$, for $1000$ cycles takes $108 \, \text{ms}$ (average of 10 runs) on a single thread of a standard 2.3 GHz Intel Core i5 processor using our in-house code, $22$ times the physical time simulated. Although the execution time is still more than one order of magnitude too large to be used in real-time control algorithms, this gap will inevitably close as faster hardware and better optimised numerical algorithms become available. Hence, the merits of studying the reliability and applicability of single-bubble models, such as the ones considered in this study, reach beyond improving our understanding of the bubble response.

\section*{Conclusions}
\label{sec:conclusions}

The primary aim of this study has been to assess the applicability and capability of commonly used single-bubble models to predict the complex response of lipid-coated microbubbles to the excitation regime frequently used for focused ultrasound applications. To this end, we have considered a SonoVue bubble with an initial radius of $1 \, \mu \text{m} \leq R_0 \leq 2 \, \mu \text{m}$, subjected to a sinusoidal excitation with frequencies ranging from $200 \, \text{kHz}$ to $1500 \, \text{kHz}$ and excitation pressure amplitudes ranging from $10 \, \text{kPa}$ to $1500 \, \text{kPa}$. Both the Rayleigh-Plesset equation and the Gilmore equation, in conjunction with the Marmottant model to represent the lipid monolayer coating, have been shown to be able to reproduce and predict key phenomena of the bubble behaviour, in particular the onset of inertial cavitation and the onset of sub- and ultraharmonic emissions. 

In addition to the validation of single-bubble models, the presented results have identified a linear regime at excitation pressure amplitudes smaller (and marginally larger) than the Blake pressure. In this linear regime, the maximum pressure at the bubble wall is linearly dependent on the excitation pressure amplitude and, likewise, the mechanical and cavitation indices. For a sufficiently large maximum bubble radius of the considered bubbles, the collapse is inertia-dominated and the maximum pressure at the bubble wall is nonlinearly dependent on the bubble radius, a quantity which can relatively easily be measured in experiments.

\revone{In summary, the very good agreement of the presented results compared to {\em in vitro} experiments under specific conditions is a further step towards a reliable quantitative prediction of the response of coated microbubbles in focused ultrasound applications.}

\section*{Acknowledgements}
\noindent 
We wish to thank both anonymous reviewers for their meticulous and insightful comments, which guided us in substantially improving the manuscript.
This research was funded by the Deutsche Forschungsgemeinschaft (DFG, German Research Foundation), grant number 441063377. 


\end{document}